\documentclass[aps,superscriptaddress,reprint,longbibliography, amsmath,amssymb,floatfix,onecolumn]{revtex4-1}
\usepackage{graphicx}
\usepackage{bm}
\usepackage{siunitx}
\usepackage{balance}
\usepackage{float}
\usepackage{color,soul}

\begin{document}
    \title{Exponential Scaling in Early-stage Agglomeration of Adhesive Particles in Turbulence}

    \author{Sheng Chen}
    \author{Shuiqing Li}
    \email[Corresponding author: ]{lishuiqing@tsinghua.edu.cn}
    \affiliation{Key Laboratory for Thermal Science and Power Engineering of Ministry of Education, Department of Energy and Power
  Engineering, Tsinghua University, Beijing 100084, China.}
    \author{Jeffrey S. Marshall}
    \affiliation{Department of Mechanical Engineering, The University of Vermont, Burlington, VT 05405, USA.}

    \date{\today}

    \begin{abstract}
We carry out direct numerical simulation together with an adhesive discrete element method calculation (DNS-DEM) to investigate agglomeration of particles in homogeneous isotropic turbulence (HIT). We report an exponential-form scaling for the size distribution of early-stage agglomerates, which is valid across a wide range of particle inertia and inter-particle adhesion values. Such scaling allows one to quantify the state of agglomeration using a single scale parameter. An agglomeration kernel is then constructed containing the information of agglomerate structures and the sticking probability. {\color{black}{An explicit relationship between the sticking probability and microscale particle properties is also proposed based on the scaling analysis of the equation for head-on collisions.}} Our results extend Smoluchowski's theory to the condition of non-coalescing solid adhesive particles and can reproduce DNS-DEM results with a simple one-dimensional simulation.
\end{abstract}

    \maketitle
\section{Introduction}
\label{intro}
Clustering of particles suspended in turbulence has been extensively studied in experiments \cite{SawPRL2008, LuPRL2010}, in simulations \cite{bec2014gravity} and by theoretical approaches \cite{Mehlig2014, balachandarAR2010}. To predict the evolution of cluster or agglomerate size, Smoluchowski's equation, built on statistical collision kernels, is one of the few theoretical tools that can be applied to large-scale systems \cite{Von1916,friedlander2000, pumir2016}. For particles in turbulence, the collision kernel is usually expressed as the production of the mean relative radial velocity and the radial distribution functions (RDFs) of particle pairs at the distance of contact. For zero-inertial particles, these two quantities can be statistically determined from those of turbulence flows \cite{SaffmanJFM1956}. In contrast, inertial particles preferentially sample certain regions of the flow due to the centrifugation effect, giving rise to higher values of both relative radial velocity and spatial concentration \cite{AbrahamsonCES1975, SundaramJFM1997, WangJFM2000,  ZhouJFM2001, gustavsson2016}. As the inertia of particles further increases, particles from different regions of the flow come together. A larger relative velocity, consequently a larger collision rate, is then observed. Such effect is termed as ``caustics" \cite{wilkinson2005caustics, gustavsson2011} or ``sling effect" \cite{falkovich2007}.


Based on these models of geometric collision kernel, Smoluchowski's theory can be then used to describe the growth of clusters assuming that colliding particles merge immediately to form new larger spherical particles. The assumption of unity coagulation efficiency is normally valid for droplets. However, it is not applicable to the agglomeration of solid {\it non-coalescing} adhesive particles. Such systems are quite ubiquitous, ranging from electrostatic agglomerators \cite{JaworekPECS2018}, flocculation during water treatment \cite{JarvisWR2005}, assemblage of preplanetary grains \cite{blum2000} to the growth of dendrites during aerosol filtration \cite{tien1977, ChenPRE2016}. The solid adhesive particles, across $1$ to $10^2$ microns, have two significant differences from Brownian nanoparticles or coalescing droplets: (1) the interparticle adhesion due to van der Waals attraction is {\it short-ranged} and relatively {\it soft} \cite{Marshall2014}. It leads to the sticking/rebound behavior of colliding particles (i.e., non-unity coagulation efficiency). (2) Formed agglomerates are usually non-spherical, whose structure will evolve due to restructuring and breakage. It has been reported that even the simplest elastic repulsion between particles considerably changes the picture of agglomeration \cite{bec2013sticky}. Constructing a kernel function that can reflect the influence of complicated inter-particle interactions is a crucial problem that has not been settled.

Solving this problem requires a fundamentally different approach - discrete element methods (DEM) - that tracks the dynamics of individual particles both while they are traveling alone through the fluid and while they are colliding with other particles \cite{Marshall2014}. To properly simulate the agglomeration, particle collisions should be resolved with a time step much smaller than the Kolmogorov time scale. Moreover, all the possible modes of particle interaction, i.e., normal impact, sliding, twisting, and rolling, should be taken into account \cite{MarshallJCP2009, DizajiPT2017}. Constructing kernel functions or stochastic agglomeration models \cite{SommerfeldIJMF2001, AlmohammedPT2016} based on data from DEM simulations then allows large-scale simulation of the agglomeration process.

In this work, we perform direct numerical simulations (DNS) to study the agglomeration of non-coalescing solid particles in homogeneous isotropic turbulence (HIT) with focus on the effect of van der Waals adhesion. A novel adhesive DEM is employed to fully resolve the translational and rotational motions of particles. We report an exponential-form scaling for the size distribution of early-stage agglomerates as $n(A)/n_0 \sim \exp(-A/\kappa)$, where $n(A)$ is the number density of agglomerates of size $A$. This exponential distribution allows one to describe the growth of agglomerates using a single scale parameter $\kappa$. Based on the simulation results, we are able to extend the Smoluchowski's theory to describe adhesion-enhanced agglomeration by introducing a turbulence agglomeration kernel depending on the fractal structure of agglomerates and an adhesion-controlled sticking probability.

\section{NUMERICAL METHOD AND SIMULATION CONDITIONS}
\label{methods}
\subsection{DNS-DEM}
\subsubsection{Fluid phase}
In our simulation, the homogeneous isotropic turbulent flow is calculated by DNS on a cubic, triply-periodic domain. A pseudospectral method with second-order Adams-Bashforth time stepping is applied to solve the continuity and momentum equations of the incompressible flow,
\begin{subequations}
\begin{align}
  \nabla \cdot {\bm u} &= 0, \label{eq:fcont} \\
  \frac{\partial {\bm u}}{\partial t} = {\bm u} \times {\bm \omega} -& \nabla\left(\frac{p}{\rho_f} + \frac{{\bm u}^2 }{2}\right)+\nu\nabla^2{\bm u}+{\bm f}_F +{\bm f}_P. \label{eq:fmom}
\end{align}
\end{subequations}
Here, ${\bm u}$ and ${\bm \omega}$ are the fluid velocity and vorticity, respectively. $p$ is the pressure, $\rho_f$ is the fluid density, $\nu$ is the kinematic viscosity. The small wavenumber forcing term ${\bm f}_F$ is used to maintain the turbulence with an approximately constant kinetic energy. As suggest in \cite{Lundgren2003, RosalesPOF2005}, we assume the forcing vector to be proportional to the fluid velocity and added to wavenumbers with magnitude $k< 5$. ${\bm f}_P$ is the particle body force, which is calculated at each Cartesian grid node $i$ using ${\bm f}_p({\bm x}_i) \!=\! - \sum_{n \!=\! 1}^N {\bm F}_n^F \delta_h \left({\bm x}_i \!-\! {\bm X}_{p,n} \right)$. Here, ${\bm x}_i$ is the location of grid node $i$, ${\bm F}_n^F$ is the fluid force on particle $n$ located at ${\bm X}_{p,n}$ and $\delta_h \left({\bm x}_i \!-\! {\bm X}_{p,n} \right)$ is a regularized delta function. {\color{black}{The influence of the particle phase on the flow phase has a non-negligible effect on the agglomeration even when the particle volume fraction $\phi<0.001$. Since we also consider interactions between particles, our simulation is four-way coupled \cite{balachandarAR2010}.

It should be noted that all the equations and variables in our simulation have been nondimensionalized by choosing typical length, velocity and mass scales that are relevant to the agglomeration of solid microparticles. The typical length scale is set as $L_0=100r_p=0.01\ {\rm m}$, where $r_p=10 \mu m$ is the particle radius. The typical velocity is $U_0=10 {\rm m/s}$ and the typical mass is $M_0=\rho_f*L_0^3=10^{-9}\ {\rm kg}$, where $\rho_f=1 {\rm kg/m^3}$ is the fluid density. The typical length scale is given by $T_0 = L_0/U_0$.Other dimensional input parameters are the fluid viscosity $\mu =1.0\times 10^{-5}\ {\rm Pa\cdot s}$, the particle density $\rho_p = 10 \sim 320\ {\rm kg/m^3}$ and the surface energy $\gamma = 0.01 \sim 5\ {\rm J/m^2}$. Hereinafter, all the variables appear in their dimensionless form and, for simplicity, we use the same notations as the dimensional variables.}}

Before the particles are added into the domain, a preliminary computation is conducted for $5000$ time steps with ${\rm d}t_F = 0.005$ (dimensionless) to allow the turbulence to reach a statistically stationary state. The turbulence kinetic energy $q$ and dissipation rate $\epsilon$ are obtained from integration of the power spectrum $E(k)$,

\begin{equation}
  \label{eq:tke}
  q = \int_0^{k_{max}} E(k) {\rm d} k,\quad \epsilon = 2 \nu \int_0^{k_{max}} k^2E(k) {\rm d}k.
\end{equation}

\subsubsection{Solid phase: adhesive discrete element method}
We use discrete element method (DEM) to model the particles' motion in turbulent flows, which solves the linear and angular momentum equations of particles
\begin{subequations}
\begin{align}
  m_i \dot{\bm v}_i = {\bm F}_i^F + {\bm F}_i^C, \label{eq:eoma} \\
  I_i \dot{\bm \Omega}_i = {\bm M}_i^F + {\bm M}_i^C. \label{eq:eomb}
\end{align}
\end{subequations}
where $m_i$ and $I_i$ are mass and moment of inertia of particle $i$ and ${\bm v}_i$ and ${\bm \Omega}_i$ are the translational velocity and the rotation rate of the particle. The forces and torques are induced by both the fluid flow (${\bm F}_i^F$ and ${\bm M}_i^F$) and the interparticle contact (${\bm F}_i^C$ and ${\bm M}_i^C$). In this work, the dominant fluid force is the Stokes drag given by
\begin{subequations}
\begin{align}
  {\bm F}^{drag} = -3\pi \mu d_p \left(\bm v - \bm u \right)f, \label{eq:fdrag} \\
  {\bm M}^{drag} = -\pi \mu d_p^3 \left({\bm \Omega} - \frac{1}{2} {\bm \omega} \right), \label{eq:mdrag}
\end{align}
\end{subequations}
{\color{black}{where ${\bm u}$, ${\bm \omega}$ and $\mu$ are velocity, vorticity and viscosity of the fluid and $\bm v$ and $d_p$ are the velocity and the diameter of particles}}. The friction factor $f$, given by \cite{di1994}, is used to correct for the crowding of particles. For particle Reynolds number in the range $0.01$ to $10^4$, $f$ can be written as
\begin{equation}
  \label{eq:di}
  f = (1-\phi)^{1-\zeta},\quad \zeta = 3.7 - 0.65 \exp\left[-\frac{1}{2}\left(1.5 - \ln Re_p \right)^2 \right].
\end{equation}
{\color{black}{The particle Reynolds number $Re_p$ is defined as $Re_p=d_p |\bm v - \bm u|/\nu$.}} In addition to the Stokes drag, we also include the Saffman and Magnus lift forces in ${\bm F}^F_i$ \cite{SaffmanJFM1965, RubinowJFM1961}.

When two particles $i$ and $j$ are in contact, the normal force $F^N$, the sliding friction $F^S$, the twisting torque $M^T$, and the rolling torque $M^R$ acting on particle $i$ from particle $j$ can be expressed as
  \begin{subequations}
      \label{eq:dem}
    \begin{align}
  F_{ij}^N & \!=\!F_{ij}^{NE}\!+\!F_{ij}^{ND} \!=\!-4F_C\left(\hat{a}^3_{ij} \!-\! \hat{a}_{ij}^{3/2} \right) \!-\! \eta_N \bm{v}_{ij}\cdot \bm{n}_{ij}, \label{eq:dem_a} \\
  F_{ij}^S &\!=\! -\mathrm{min}\left[ k_T\int_{t_0}^t \bm{v}_{ij}(\tau)\cdot \bm{\xi}_S \mathrm{d}\tau \!+\!\eta_T\bm{v}_{ij}\cdot \bm{\xi}_S,\              F_{ij,crit}^S \right],        \label{eq:dem_b} \\
   M_{ij}^T &\!=\! -\mathrm{min}\left[ \frac{k_Ta^2}{2}\int_{t_0}^t \bm{\Omega}_{ij}^T(\tau)\cdot \bm{n}_{ij} \mathrm{d}\tau \!+\! \frac{\eta_Ta^2}{2}\bm{\Omega}_{ij}^T\cdot \bm{n}_{ij},\ M_{ij,crit}^T \right], \label{eq:dem_c} \\
  M_{ij}^R &\!=\! -\mathrm{min}\left[ 4F_C\hat{a}_{ij}^{3/2}\int_{t_0}^t \bm{v}_{ij}^L(\tau)\cdot \bm{t}_R \mathrm{d}\tau \!+\!\eta_R \bm{v}_{ij}^L\cdot \bm{t}_R,\ M_{ij,crit}^R \right]. \label{eq:dem_d}
\end{align}
\end{subequations}
The normal force $F_{ij}^N$ contains an elastic term $F_{ij}^{NE}$ derived from the JKR (Johnson-Kendall-Roberts) contact theory. $F^{NE}$ combines the effects of van der Waals attraction and the elastic deformation and its scale is set by the critical pull-off force, $F_C = 3\pi R_{ij}\gamma$, where $R_{ij} = (r_{p,i}^{-1} +r_{p,j}^{-1})^{-1}$ is the reduced particle radius and $\gamma$ is the surface energy of the particle. {\color{black}{The dimensionless variable $\hat{a}_{ij}$ is calculated by normalizing the radius of the contact region $a_{ij}$ by its value at the zero-load equilibrium state $a_{ij,0}$, expressed as $a_{ij,0} = (9\pi\gamma R_{ij}^2/E_{ij})^{1/3}$ \citep{MarshallJCP2009}, where $E_{ij}$ is the effective elastic modulus.}} In DEM, $\hat{a}_{ij}$ is calculated inversely from the normal particle overlap, $\delta$, through
\begin{equation}
  \label{eq:delta_a}
  \frac{\delta}{\delta_C} = 6^{\frac{1}{3}}\left[2(\hat{a}_{ij})^2 - \frac{4}{3}(\hat{a}_{ij})^{\frac{1}{2}} \right],
\end{equation}
where $\delta_C = a^2_{ij,0}/(2(6)^{1/3}R_{ij})$ is the critical overlap. {\color{black}{The bond between two contacting particles will break when $\delta<-\delta_C$. The sliding friction $F^S$, twisting torque $M^T$, and rolling torque $M^R$ (Eq. (\ref{eq:dem_b}) - (\ref{eq:dem_d})) are all calculated using spring-dashpot-slider models, where $\bm{v}_{ij}\cdot\bm{\xi}_S$, $\bm{\Omega}_{ij}^T$, and $\bm{v}_{ij}^L$ are the relative sliding, twisting, and rolling velocities. $k_T$ in Eq. (\ref{eq:dem}) is the tangential stiffness. The second term of Eq. (\ref{eq:dem_a}-\ref{eq:dem_d}) are the viscoelastic damping forces, which are proportional to the rate of motions in each of the respective directions, and $\eta_N$, $\eta_T$ and $\eta_R$ are the dissipation coefficients for relative compression, sliding and rolling motions. The normal dissipation coefficient $\eta_N$ is calculated as $\eta_N = 2\alpha \sqrt{m_{ij}a_{ij}E_{ij}/3}$, where $m_{ij} = (m_i + m_j)^{-1}$ is the effective mass of two colliding particles with mass $m_i$ and $m_j$. For details, see \cite{MarshallJCP2009, chen2019fast}.}}

When these resistances reach their critical limits, $F_{ij,crit}^S$, $M_{ij,crit}^T$ or $M_{ij,crit}^R$, a particle will irreversibly slide, twist or roll relative to its neighboring particle. The critical limits are expressed as \cite{MarshallJCP2009}:
\begin{subequations}
    \label{eqcrit}
\begin{align}
  F_{ij,crit}^S &= \mu F_C \left|4\left(\hat{a}_{ij}^3 - \hat{a}^{3/2}_{ij}\right) + 2 \right|, \label{eqcrit_a} \\
  M_{ij,crit}^T &= \frac{3\pi a_{ij} F_{ij,crit}^S} {16}, \label{eqcrit_b}\\
  M_{ij,crit}^R &= 4F_C\hat{a}_{ij}^{3/2} \theta_{crit}R_{ij}. \label{eqcrit_c}
\end{align}
\end{subequations}
Here $\mu (= 0.3)$ is the friction coefficient and $\theta_{crit} (= 0.01)$ is the critical rolling angle. We set these values according to experimental measurements \cite{SumerJAST2008}. {\color{black}{The adhesive DEM has been validated by a series of experimental measurements. The details of these validations and the determination of the value of parameters in DEM can be found in \cite{YangPT2013, chen2019fast}.}}

{\color{black}{The scales of the elastic term $F_{ij}^{NE}$ in Eq. (\ref{eq:dem_a}) and the critical force and torques in Eq. (\ref{eqcrit}) are all in proportion to the surface energy $\gamma$, which is the work required to separate two touching surfaces per unit area. An adhesion parameter $Ad$, which is defined as the ratio between $\gamma$ and the kinetic energy of particles (per unit area), can be used to quantify the effect of adhesion. $Ad$ is expressed as \cite{LiJAS2007, chen2019fast, Marshall2014}
\begin{equation}
    \label{ad_define}
    Ad = \frac{\gamma}{\rho_p U^2 r_p}
\end{equation}
In this equation, $U$ is the characteristic velocity scale of particles. For particles transported in turbulence, we simply set $U$ equal to the root-mean-square turbulent fluctuation velocity $u^{\prime}$. An alternative choice of the velocity scale is discussed in Sec. \ref{sec_theta}. For large values of the adhesion parameter, particles tend to stick together upon collision, forming particle agglomerates. In contrast, colliding particles tend to rebound from each other when $Ad$ is small. $Ad$ has been successfully used to estimate the critical sticking velocity of two colliding particles \cite{ChenPT2015} and predict the packing structure of adhesive particles \cite{LiuSM2015, LiuSM2017, ChenPRE2016}.}}

It is known that the fluid squeeze-film between particles near contact significantly reduces the approach velocity and further influences the collision and agglomeration process. In this work, viscous damping force derived from the classical lubrication theory is also included, given by
\begin{equation}
  \label{eq:lubrication}
  F_l = -\frac{3\pi\mu r_p^2}{2h}\frac{{\rm d}h}{{\rm d}t}.
\end{equation}
$F_l$ is initiated at surface separation distance $h \!=\! 0.01r_p$ and a minimum value $h\!=\! 2\times10^{-4}r_p$ is set at the instant of particle contact according to experiments \cite{Marshall2011,YangPOF2006}.

\subsubsection{Multiple-time step framework}
Our DNS-DEM computational framework is designed with multiple-time steps \cite{LiJAS2007, MarshallJCP2009, Marshall2014}. The flow field is updated using a fluid time step ${\rm d}t_F = 0.005$. To correctly identify inter-particle collisions, a smaller particle convective time step ${\rm d}t_P = 2.5\times 10^{-4}$ is adopted to update the force, velocity, and position of particles that do not collide with other particles. Such a small ${\rm d}t_p$ ensures that the distance each particle travels during a time step is only a small fraction of the particle or the grid size. In addition, we build a local list at each fluid step to record the neighboring particles that each particle may collide as it is advected over a fluid time step. Once a particle is found to collide with other particles during a particle time step, we then recover its information (i.e., its force, velocity, and position) to the start of this particle time step and instead advect it using a collision time step ${\rm d}t_C = 6.25\times10^{-6}$. The value of ${\rm d}t_C$ is small enough to resolve the rapid variation of contacting forces, velocity, and position of the particles.

\subsection{Simulation conditions}
\begin{figure}
  \includegraphics[width = 10.5 cm]{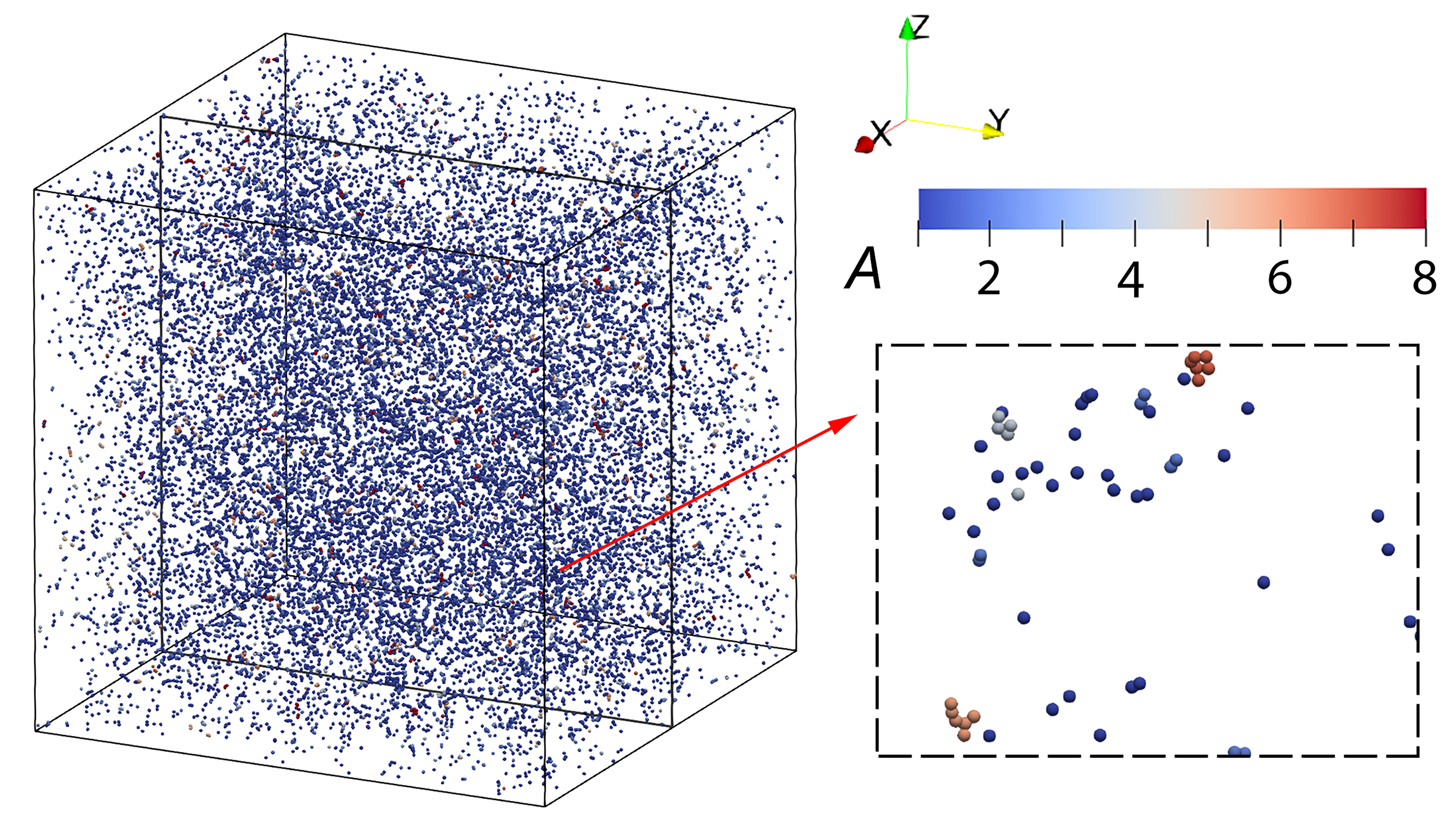}
  \centering
  \caption{\label{fig:snapshot} Snapshot of the simulated system at $t = 20$. The enlarged view from the middle slice ($x=0$) shows agglomerates and their size $A$ (defined as the number of primary particles contained in the agglomerate, indicated by the colorcode).}
\end{figure}

The system studied in this work is illustrated in Fig. \ref{fig:snapshot}. We consider $N \!=\! 4\times10^4$ non-Brownian solid particles suspended in the homogeneous isotropic turbulent flow {\color{black}{in the absence of gravity}}. The triply periodic computational domain has a dimension of $(2\pi)^3$ with $128^3$ grid points. The Taylor Reynolds number is fixed as $Re_{\lambda} = 93.0$ in this work. Similar values of $Re_{\lambda}$ have also been used in previous studies involving particle-laden flows \cite{WangJFM2000,fayed2013, JinJOE2017}. By setting this value of $Re_{\lambda}$, we can easily compare our results with those in literature.  Other dimensionless flow parameters, including the fluctuating velocity $u^{\prime}$, the dissipation rate $\epsilon$, the kinematic viscosity $\nu$, Kolmogorov length $\eta$, Kolmogorov time $\tau_k$, and the large-eddy turnover time $T_e$, are listed in Table \ref{tab:turb}.

The particle radius is fixed as $r_p \!=\! 0.01$. We choose the value of particle radius so that the particle size and the Kolmogorov length scale are comparable. We choose this relatively large value of particle size to increase the collision rates, which helps ensure good statistics on agglomeration within a feasible computing time. The particle volume concentration is $\phi=\frac{4N\pi r_p^3}{3(2\pi)^3} = 6.7 \times 10^{-4}$, which is small so that the system can be regarded as a dilute system. {\color{black}{The fluid density $\rho_f$ is set as 1 (non-dimensional), and five different values (10, 40, 80, 160 and 320) are used as particle density $\rho_p$ to achieve different values of particle response time. We have neglected the influence of gravity in the present study since it does not play an important role in the agglomeration of particles with radius less than $40 {\it \mu m}$ \cite{pruppacher2012microphysics}. For detailed discussions on the effect of gravity on collision rate for large particles (with size above $40 {\it \mu m}$), we refer to \cite{onishi2009influence,ireland2016gravity}}}

One of the most important parameters governing the agglomeration is the Kolmogorov-scale Stokes number, $St_k = \tau_p/\tau_k$, where $\tau_p \!=\! m/(6\pi r_p \mu)$ is the particle response time and $\tau_{k} \!=\! (\nu /\epsilon)^{1/2}$ is the Kolmogorov time. In the classical theory of turbulent collision of nonadhesive particles, $St_k$ significantly influences the value of the collision kernel. In the presence of adhesion, the adhesion parameter $Ad = \gamma/(\rho_p u^{\prime 2}r_p)$ is used to quantify the adhesion effect \cite{Marshall2014}. The particle surface energy $\gamma$ can be determined according to experimental measurements \cite{SumerJAST2008, krijt2013energy} or calculated from the Hamaker coefficients of the materials \cite{Marshall2014}. {\color{black}{In this work, we systematically vary $Ad$ (by varying $\gamma$) in a wide range at five different $St_k$ values (0.72, 2.9, 5.8, 12 and 23) to show the effect of adhesion on the agglomeration.}}

\begin{table}
    \setlength{\tabcolsep}{7pt}
    \centering
    \caption{\label{tab:turb} Dimensionless parameters of the fluid turbulence, including the fluctuating velocity $u^{\prime}$, the dissipation rate $\epsilon$, the kinematic viscosity $\nu$, Taylor-microscale Reynolds number $Re_{\lambda}$, Kolmogorov length $\eta$, Kolmogorov time $\tau_k$, and the large-eddy turnover time $T_e$.}
    \begin{tabular}{*{7}{c}}
    \hline
    \hline
  $u^{\prime}$   &   $\epsilon$    &  $\nu$   &  $Re_{\lambda}$  &  $\eta$  & $\tau_k$ &  $T_e$\\
    \hline

  0.28   &   0.0105             &  0.001  &  93.0           &  0.0175  &  0.31    & 7.4\\
    \hline
    \hline
    \end{tabular}
\end{table}

\subsection{Smoluchowski's theory}
Before showing the DNS-DEM results, we introduce the Smoluchowski coagulation equation and discuss how to apply the theory to the agglomeration of non-coalescing adhesive particles. In Smoluchowski's theory, the growth of agglomerates can be described using the population balance equation (PBE) \cite{Von1916}
\begin{equation}
  \label{eq:Smoluchowski}
  \dot{n}(A)\! =\! \!\frac{1}{2}\! \!\sum_{i+j = A} \! \Gamma(i,j)n(i)n(j)\!-\! \!n(A)\!\sum_{i = 1}^{\infty} \Gamma(i,A)n(i),
\end{equation}
where $\Gamma(i,j)$ is the averaged rate constant (kernel) for agglomerates of size $i$ colliding with agglomerates of size $j$ and should reflect all the factors affecting agglomeration. It is defined as $\Gamma(i,j) \equiv \dot{n}_{c,ij}/(n(i)n(j))$ with $\dot{n}_{c,ij}$ being the collision rate per unit volume and $n(i)$ being the average number concentration of size group $i$. The first term on the right-hand side of Eq. (\ref{eq:Smoluchowski}) is the source term that accounts for the rate at which agglomerates of size $A$ are created. The second term is a sink that describes agglomerate disappearance due to its coalescence with other agglomerates.

PBE can be readily used to predict the growth of droplets in clouds with an underlying assumption - colliding particles coalesce instantaneously to form larger particles \cite{chen2018turbulence}. Therefore, the growth rate of agglomerates is equivalent to the collision rate. The collision between adhesive non-coalescing microparticles, however, does not ensure the growth of an agglomerate. Both sticking and rebound could happen as a result of the competition between the particles' kinetic energy and the surface energy. Thus, it is natural to introduce a sticking probability, $\Theta$, defined as the ratio of the number of collisions that lead to agglomeration to the total number of collisions. We then have an {\it agglomeration kernel}, which reads
\begin{equation}
  \label{eq:sticking}
 \Gamma_a(i,j) = \Theta \Gamma(i,j),\ \forall i,j.
\end{equation}
The sticking probability has a minimum value $0$ for non-adhesive particle systems and a maximum value $1$, corresponding to the hit-and-stick case in conventional PBE simulations. We can then simulate the agglomeration with different adhesion level, by simply replacing $\Gamma(i,j)$ in Eq. (\ref{eq:Smoluchowski}) by $\Gamma_a(i,j)$. We will show below that such simple modification can well reproduce DNS-DEM results in a statistical manner.

The structure of agglomerates is another crucial factor affecting the agglomeration rate. For non-coalescing adhesive particles, the formed agglomerates usually have fractal structures, which distinguishes our system from those of droplets \cite{dizaji2016accelerated, DizajiPT2017}. In systems involving Brownian nanoparticles, theoretical collision kernels can be extended to fractal agglomerates when substituting the particle radius with the radius of effective collision spheres (ECSs) for an agglomerate \cite{JiangEST1991, flesch1999laminar}. We will show below that the idea of the effective radius can also be applied to non-Brownian inertial particles.

\section{Results and Discussions}
\label{results}
\subsection{Collision rate, agglomerate size, and structure}
\begin{figure}
  \includegraphics[width = 11.5 cm]{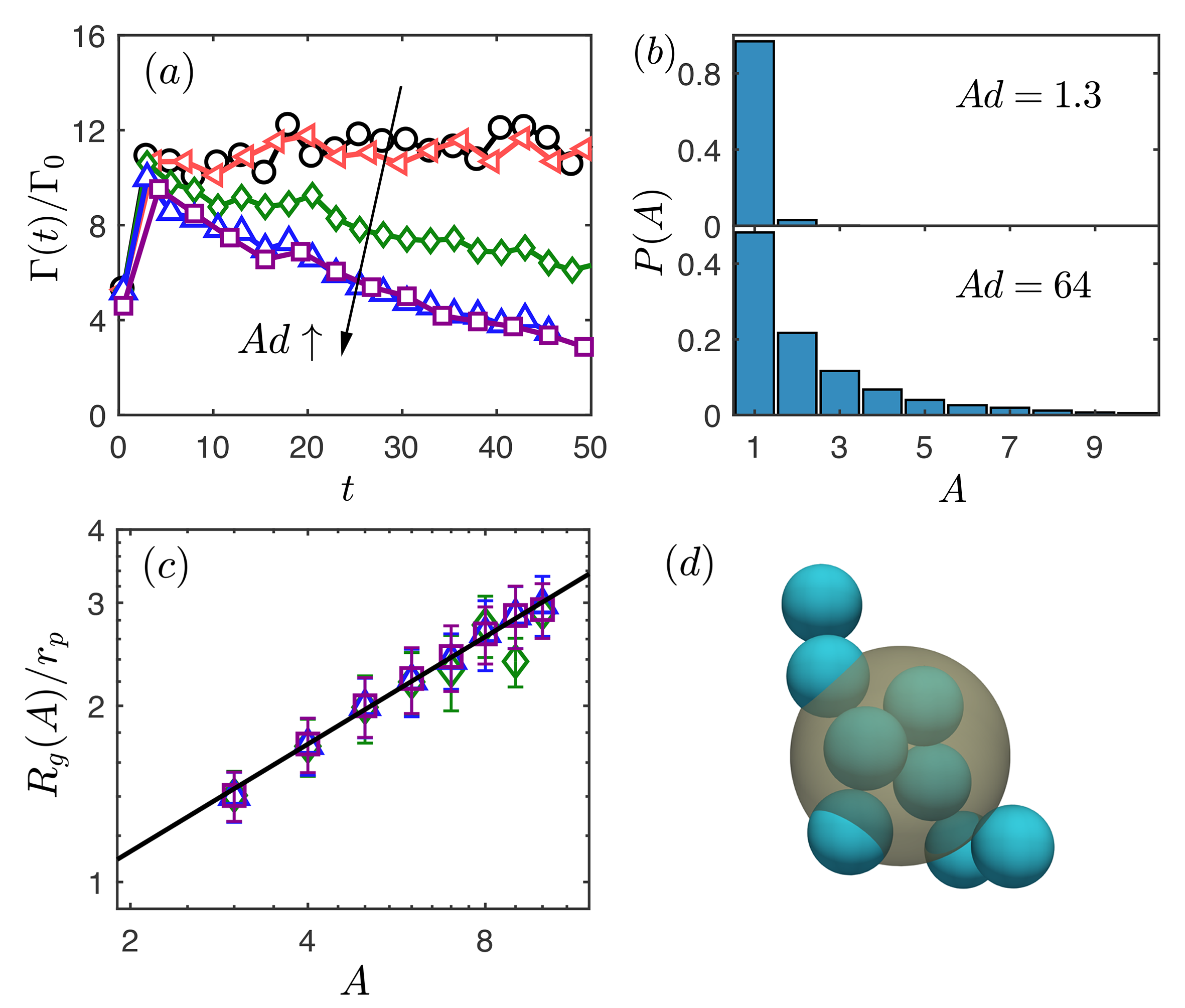}
  \centering
  \caption{\label{fig:gyr} (a) Temporal evolution of the collision kernel $\Gamma(t)/\Gamma_0$ for cases with $St_k = 5.8$ and $Ad = 0.013$ (circles), $1.3$ (left-pointing triangles), $13$ (diamonds), $64$ (upward triangles), and $128$ (squares). (b) Fraction of particles, $P(A)$, contained in agglomerates of size $A$ at $t=15$ for $Ad = 1.3$ and $64$. (c) Gyration radius of agglomerates $R_g(A)/r_p$ as a function of agglomerate size $A$ at $t=15$ for the cases with $St_k = 5.8$ and $Ad = 13$ (diamonds), $64$ (triangles), and $128$ (squares). The solid line shows Eq. (\ref{eq:rgy}) with $k = 1.64$ and $D_f = 1.64$. (d) An agglomerate produced in the simulation with $St = 5.8$ and $Ad = 64$ with its equivalent sphere with radius of gyration (shaded region).}
\end{figure}

We first measure the temporal evolution of the collision kernel in a system with $St_k = 5.8$ and $Ad$ varying from $0.013$ to $128$. To show the adhesion effect, here we simply regard the system as a monodisperse system and count the collisions between every primary particle. The collision kernel is then calculated as $\Gamma = 2\dot{n}_{c}/n_0^2$, where $n_0$ is the number density of primary particles. The temporal evolution of the collision kernel $\Gamma(t)$, normalized by the collision kernel for zero-inertia particles $\Gamma_0 = (8\pi\epsilon/15\nu)^{1/2}(2r_p)^3$ \cite{SaffmanJFM1956}, is shown in Fig. \ref{fig:gyr} (a). When the adhesion is extremely weak ($Ad = 0.013$ and $1.3$), the collision kernel rapidly reaches a statistically steady state with $\Gamma(t)/\Gamma_0 = 11.1$. {\color{black}{This value is quite close to the previous DNS results for nonadhesive particles with the same inertia \cite{WangJFM2000}}}. As $Ad$ increases, the collision kernel is significantly reduced and the system is pushed away from equilibrium. Since adhesion number only affects the interaction between contacting particles, we attribute these phenomena to adhesion-enhanced agglomeration. {\color{black} {When $Ad$ is larger than $64$, further increase of $Ad$ does not change the curve of collision kernel. Because in this strong-adhesion limit, the sticking probability for colliding particles is essentially unit and every collision event will lead to agglomeration. The overall collision kernel is determined by the size distribution of the agglomerates in the system, which is mainly determined by the turbulent transport and is insensitive to the adhesion in this large $Ad$ limit.}}

In Fig. \ref{fig:gyr} (b), the agglomeration at $t\!=\! 15$ is clearly displayed in the form of the fraction of particles $P(A)$ contained in an agglomerate of size $A$. {\color{black}{The agglomerate size $A$ is defined as the number of primary particles contained in that agglomerate.}} For small $Ad$, most particles remain as singlets ($A=1$) and only a small number of particles ($\sim 4\%$) are contained in agglomerates of size $A\ge 2$. In contrast, cases with large $Ad$ yield a considerable number of agglomerates with size $A$ up to $20$.

To model the agglomeration process in the framework of Smoluchowski's equation, a measure of agglomerate structure in the form of the equivalent sphere is necessary. One such quantity is the radius of gyration, defined for an agglomerate with $3$ or more primary particles ($A \ge 3$) by $R_g(A) = (\Sigma_1^{A}|{\bm X}_i - \bar{\bm X}_i|^2/A)^{1/2}$, where ${\bm X}_i$ denotes the position of $i^{\rm th}$ particle within the agglomerate and $\bar{\bm X}_i$ is the centre of mass of the agglomerate.
For agglomerates with 2 primary particles, we use the explicit expression $R_g(2) = \sqrt{1.6}r_p$ suggested by \cite{WaldnerPT2005}.

In Fig. \ref{fig:gyr} (d), we show an agglomerate generated from DNS-DEM simulation and its equivalent sphere with the radius of gyration. We calculate $R_g$ for all the agglomerates produced in the simulations in Fig. \ref{fig:gyr}(a) at $t\!=\! 15$ and plot the ratio $R_g/r_p$ as a function of agglomerate size $A$ in {\color{black}{Fig. \ref{fig:gyr} (c)}} (large size agglomerates with $A>12$ only contain $~0.2\%$ particles thus are neglected here). The results fall onto a power-law curve
\begin{equation}
    \label{eq:rgy}
    \frac{R_g(A)}{r_p} = \left(\frac{A}{k} \right)^{\frac{1}{D_f}},\ {\rm for}\ A>2,
\end{equation}
with the factor $k = 1.64$ and the fractal dimension $D_f = 1.64$. The $D_f$ value measured here is consistent with experimental measurements of {\color{black}{Waldner et al. \cite{WaldnerPT2005}, who measured the radius of gyration for early-stage agglomerates formed in a stirred tank using small angle static light scattering \cite{WaldnerPT2005}. The value of fractal dimension fitted from experimental results is $D_f=1.7 \pm 0.1$, which is consistent with results of our simulations. Selomulya et al. adopted the same experimental technique to measure the shear-induced agglomeration of latex particles and reported values of $D_f$ between $1.7$ and $2.1$ \cite{selomulya2001evidence}. Their results are close to but slightly larger than the values of $D_f$ measured in our DNS-DEM results. The possible reason for the deviation is that Selomulya et al. assumed the factor $k$ to be $1.01$ in their measurements.  Such a small value of $k$ may give $D_f$ that is larger than the actual value.}} It should be noted that we focus on the agglomeration at early-stage in the current study, when the restructuring and breakage of agglomerates are normally not involved \cite{WaldnerPT2005}. These phenomena will lead to a variation of factor $k$ and the fractal dimension $D_f$ \cite{LiuPT2018}, which is left for future work.

\subsection{Effect of Stokes number}
\begin{figure*}[h]
    \includegraphics[width = 16 cm]{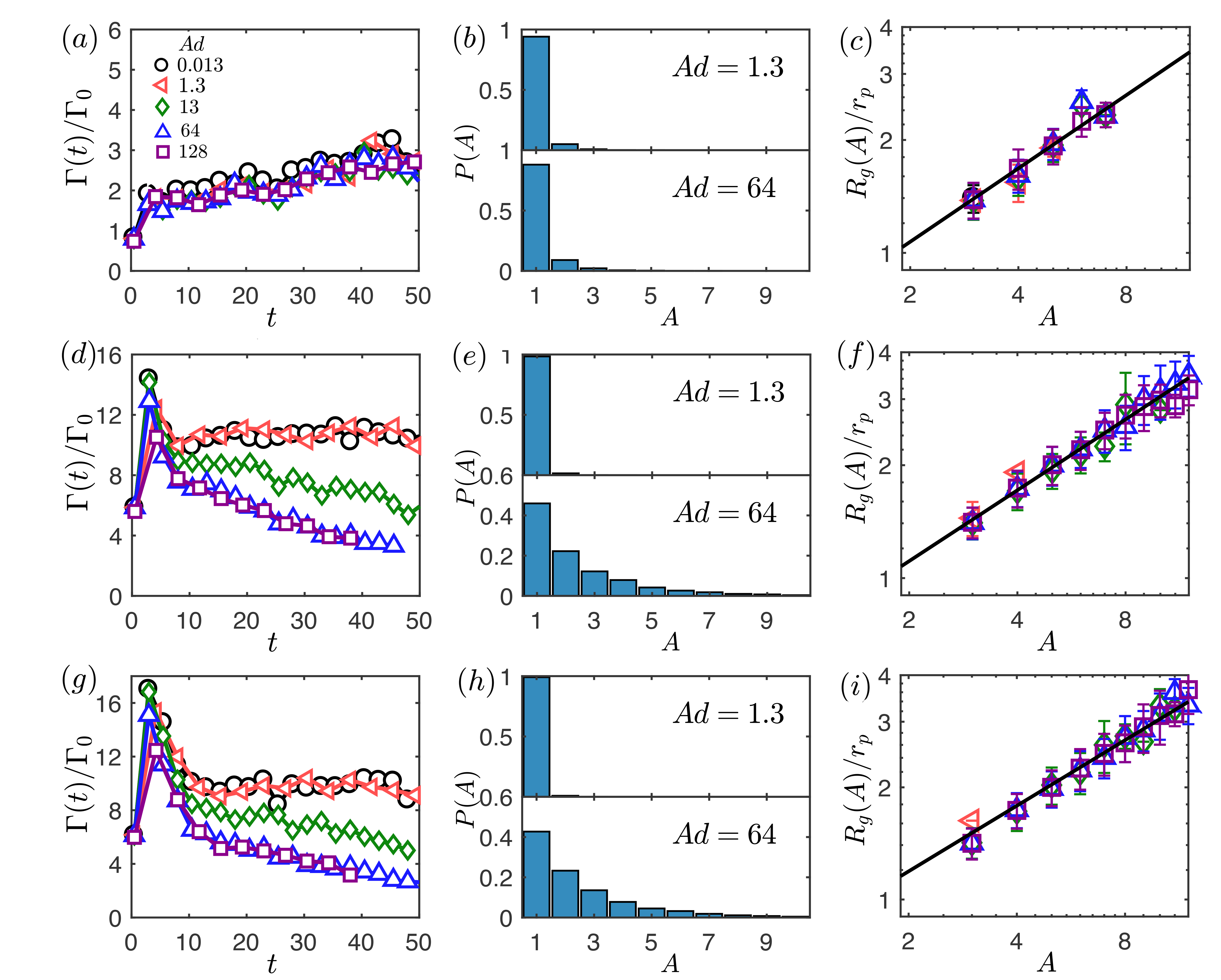}
    \centering
    \caption{\label{fig:st}{\color{black}{Left (panels (a), (d) and (g)): Temporal evolution of the collision kernel $\Gamma(t)/\Gamma_0$.  Middle (panels (b), (e) and (h)): Fraction of particles, $P(A)$, contained in agglomerates of size $A$ at $t=15$ for $Ad = 1.3$ and $64$. Right (panels (c), (f) and (i)): Gyration radius of agglomerates $R_g(A)/r_p$ as a function of agglomerate size $A$ at $t = 15$. The solid lines in (c), (f), and (i) are fits to Eq. (\ref{eq:rgy}) with (c) $k = 1.80$ and $D_f = 1.54$, (f) $k = 1.70$, $D_f = 1.60$, and (i) $k = 1.49$, $D_f = 1.71$. Different rows stand for results for different $St_k$. }}}
\end{figure*}

The temporal evolution of the collision kernel $\Gamma(t)/\Gamma_0$, the fraction of particles, $P(A)$, contained in agglomerates of size $A$, and the gyration radius of agglomerates $R_g(A)/r_p$ for cases with different Stokes number $St_k$ and adhesion parameter $Ad$ are plotted in Fig.~\ref{fig:st}. {\color{black}{For particles with small inertia ($St_k = 0.72$), the increase of adhesion parameter only has a limited effect on the temporal evolution of the collision kernel (Fig. \ref{fig:st}(a)). Moreover, there is no obvious statistical steady state for the system with $St_k = 0.72$. The reason is that the lubrication force between particles near contact significantly reduces the collision rate for particles with small inertia \cite{Marshall2011} and the collision rate is too small to form a considerable number of agglomerates even if the adhesion is strong.}} The system thus behaves as a monodisperse system. This is further displayed in the form of the fraction of particles $P(A)$ contained in an agglomerate of size $A$ (Fig. \ref{fig:st}(b)). In both strong and weak adhesion cases, most particles remain as singlets.

{\color{black}{For particles with higher Stokes number, $St_k = 12$ or $23$, similar results are observed as those for $St_k = 5.8$ in Fig.\ref{fig:gyr}. In both cases, a statistical steady state can be identified in the temporal evolution of the collision kernel $\Gamma(t)/\Gamma_0$ at the small $Ad$ limit (Fig. \ref{fig:st}(d) and (g)). When $Ad>64$, further increase of $Ad$ does not change the $\Gamma(t)/\Gamma_0 - t$ curves. The results once again confirm the existence of the strong adhesion limit. In this limit, one can simply adopt the hit-and-stick assumption - two particles will stick together once there is a contact between them - to simulate the agglomeration without performing DEM calculations. In Fig. \ref{fig:st}(e) and (d), we observe similar results as those for $St_k = 5.8$ in Fig.\ref{fig:gyr}(b).}}

{\color{black}{For all the three values of $St_k$, the radius of gyration for agglomerates of different size can be well described using the power-law function in Eq. (\ref{eq:rgy}) (see Fig. \ref{fig:st}(c), (f) and (i)). For a given $St_k$, the factor $k$ and fractal dimension $D_f$ are insensitive to the value of adhesion parameter $Ad$. It suggests that the interparticle adhesion strongly affects the growth rate of early-stage agglomerates but have no obvious impact on their structures. Interestingly, as we mentioned in the previous subsection, the agglomerates formed in different experimental conditions also have similar values of $D_f$, which further implies that the influences of flow conditions and interparticle adhesion on the structure of agglomerates may be significant only if the size of agglomerates is sufficiently large \cite{LiuPT2018}.}}

\subsection{Exponential scaling of early stage agglomerate size}
\begin{figure}
  \includegraphics[width = 16.5 cm]{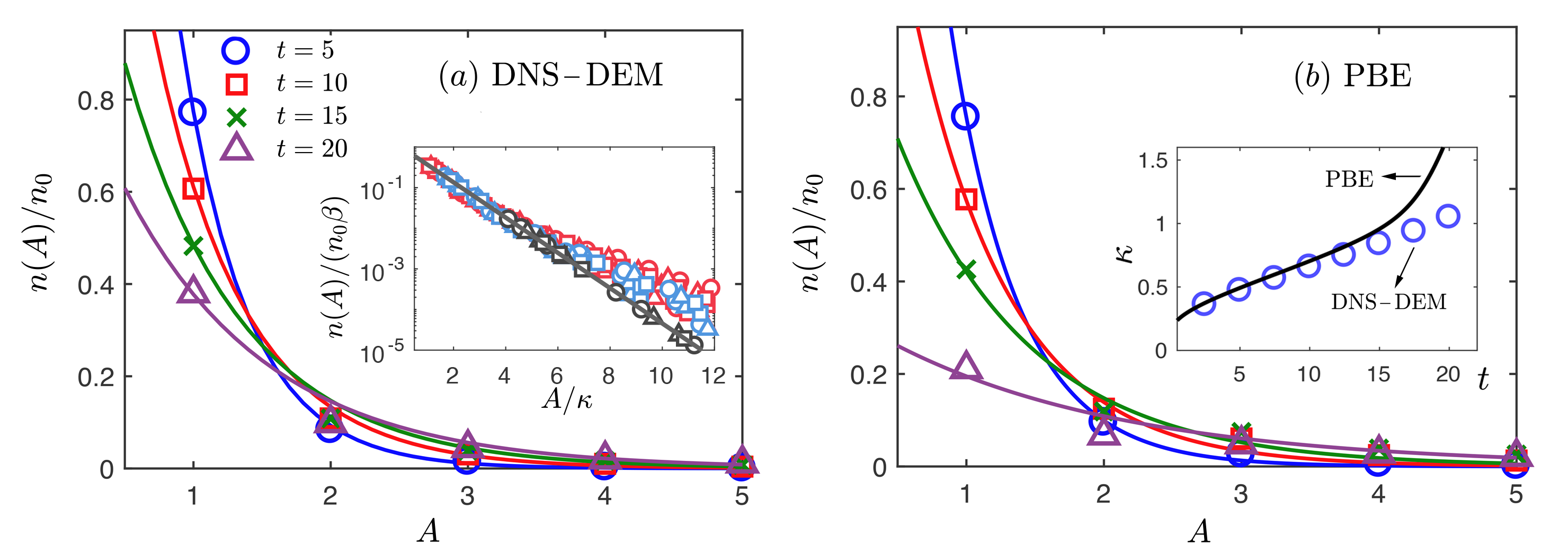}
  \centering
  \caption{\label{fig:exponential} (a) Scaled number density $n(A)/n_0$ of agglomerates of size $A$ for the case with $St_k = 5.8$ and $Ad = 64$ at $t = 5$ (circles), $10$ (squares), $15$ (exes), and $20$ (triangles). The solid lines are fits to Eq. (\ref{eq:exponential}). Inset: scaled number density $n(A)/(n_0\beta)$ as a function of $A/\kappa$ for $St_k = 5.8$ (circles), $12$ (triangles), and $23$ (squares). For each $St_k$, results are shown for $Ad = 1.3$ (black), $13$ (blue), and $64$ (red), at $t = 5$, $10$, $15$. The solid line is the exponential scaling $y = \exp (-x)$. (b) $n(A)/n_0$ v.s. $A$ calculated from population balance equations. Legends are the same as panel (a). In the inset of (b), we show the temporal evolution of the scale parameter $\kappa$ from DNS-DEM result (circles) and from PBE (solid line).}
\end{figure}

Fig. \ref{fig:exponential}(a) shows the distributions of number density of agglomerates as a function of size $A$ at early-stage ($t\le 20$). These distributions, when scaled by the initial number density of primary particles $n_0$, follow an exponential equation (solid lines in Fig. \ref{fig:exponential}(a))
\begin{equation}
\label{eq:exponential}
  \frac{n(A)}{n_0} = \beta \exp \left( - \frac{A}{\kappa} \right),
\end{equation}
with the coefficients $\beta$ and $\kappa$ depending on time. Based on the conservation of the total number of primary particles, $\Sigma_1^{\infty} A*n(A)= n_0 $, the prefactor $\beta$ can be expressed as $\beta(\kappa) = 2 \cosh{(\kappa^{-1})} - 2$. Therefore, the size distribution of early-stage agglomerates is determined by a single scale parameter $\kappa$, {\color{black}{which gives a typical value of the size of agglomerates. A larger value of $\kappa$ means that there are more particles contained in agglomerates with larger size and the growth of early-stage agglomerates can be characterized by the increase of $\kappa$.}} In the inset of Fig. \ref{fig:exponential}(a), the number density distributions for cases with $St_k = 5.8$, $12$, and $23$ and $Ad = 1.3$, $13$, and $64$ are plotted in a rescaled form, $n(A)/(n_0\beta) \sim A/\kappa$. Except for the deviation in tail caused by agglomerates with $n(A)/n_0<0.3\%$, the results center around the curve $y = \exp (-x)$, suggesting that the exponential scaling for early-stage agglomeration is valid for inertial particles across a wide range of adhesion force magnitudes.

{\color{black}{A comparison between the exponential distribution and the well-known self-preserving size distribution for Brownian nanoparticles  \cite{Friedlander1966, Vemury1995, eggersdorfer2014} would be of interest. If the collision kernels are homogeneous functions of the volume of colliding particles and the degree of homogeneity smaller than unity, the particle size distribution will reach a self-preserving shape (normally bell-shaped). In that case, tracking the evolution of the mean agglomerate size is sufficient to describe the growth of agglomerates. Although, both the exponential distribution in Eq. (\ref{eq:exponential}) and the self-preserving size distribution are single-parameter distributions, there is a fundamental difference between them. The exponential distribution describes the transition behavior at the early-stage of the agglomeration when most particles remain as singlets and is no longer valid when there is a considerable number of large agglomerates.  In contrast, the self-preserving size distribution is an asymptotic limit which is invariant with time.}}

Now we introduce how to construct the agglomeration kernel that can be applied to Smoluchowski's theory based on DNS-DEM results.  We first look at the strong adhesion case by assuming that particles will stick together upon collisions (i.e., $\Theta = 1$) and then show how adhesion influences the sticking probability. For spherical particles, $\Gamma(i,j)$ is given by
\begin{equation}
    \label{eq:kernel}
    \Gamma(i,j) = 2\pi R_{ij}^2\langle |w_r| \rangle g(R_{ij}),
\end{equation}
where $R_{ij} = r_{p,i} + r_{p,j}$ is the collision radius, $\langle |w_r| \rangle$ is the average radial relative velocity and $g(R_{ij})$ is the radial distribution function at contact. Explicit expressions of these quantities are summarized in \cite{ZhouJFM2001}. Since turbulence parameters are fixed here, $R_{ij}$, $\langle w_r \rangle$ and $g(R_{ij})$ are determined by particle size and $St_k$. For collisions between agglomerates, we simply use the radius of gyration in Eq. (\ref{eq:rgy}) with known values of $k$ and $D_f$ instead of the particle radius $r_p$ to calculate all the quantities in Eq. (\ref{eq:kernel}) \cite{Jullien1989,JiangEST1991, Eimelech1998}. {\color{black}{For instance, the collision radius for an agglomerate with $i$ primary particles and that with $j$ primary particles is calculated as $R_{ij} = R_g(i) + R_g(j)$. The gyration radius $R_g(i)$ is given by Eq. (\ref{eq:rgy}) when $i>2$ and $R_g(i)= r_p$ and $\sqrt{1.6}r_p$ for $i=1$ and $i=2$, respectively. Given the initial conditions, $n(1) = n_0$ and $n(i) = 0$ for $i>1$, PBEs in Eq. (\ref{eq:Smoluchowski}) are numerically integrated using a sufficiently small time step with the agglomerate size truncated at $i_{C} = 50$ (i.e., assuming $n(i) =0$ for $i>50$). As a result, we can get the evolution of the number density $n(i)$ for each size group. PBE calculations are much faster than DNS-DEM, since PBEs only solve for the number density $n(i)$ at each time step rather than resolve the motion of every particle.

We plot the scaled number density $n(A)/n_0$ calculated from PBE in Fig. \ref{fig:exponential}(b). It is shown that results from PBE well reproduce the results of DNS-DEM in Fig.\ref{fig:exponential}(a) when $t \le 15$. We then fit the scaled distribution $n(A)/n_0$ using the Eq. (\ref{eq:exponential}) at each $t$ and get the evolution of the scale parameter $\kappa$, which is in good agreement with the DNS results when $t \le 15$ (see the inset of Fig. \ref{fig:exponential}(b)). It indicates that the kernel $\Gamma(i,j)$ constructed in the form of gyration radius readily reflects the effect of the fractal structure of agglomerates on the agglomeration.}} At $t = 20$, the distribution of $n(A)/n_0$ from PBE still follows the exponential form, however, a non-negligible deviation between PBE results and those from DNS-DEM is observed. Such deviation may be attributed to two reasons. First, $\Gamma(i,j)$ does not contain information of breakage or rearrangement, which is expected to be significant for large-size agglomerates \cite{DizajiPT2017}. Moreover, statistics may also get worse when the total number of agglomerates $\Sigma_1^{\infty} n(A)$ reduces.

\subsection{Effect of adhesion on growth of agglomerates}
\label{sec_ad}
When the adhesion is relatively weak, a collision between two particles or agglomerates does not ensure the formation of a larger agglomerate. {\color{black}{The adhesive DEM approach can capture the effect of adhesion on the agglomeration without any additional models. However, when designing large-scale devices, one does not need to know the information of every single particle, instead, knowing the size distribution is enough. In those cases, solving the population balance equations is more feasible. Therefore, it is of significance to check if the complicated effect of particle-particle contacting interactions on the growth kinetics of agglomerates can be captured by the sticking probability $\Theta$ (given in Eq. (\ref{eq:sticking})).}}

\begin{figure}
  \includegraphics[width = 16.5 cm]{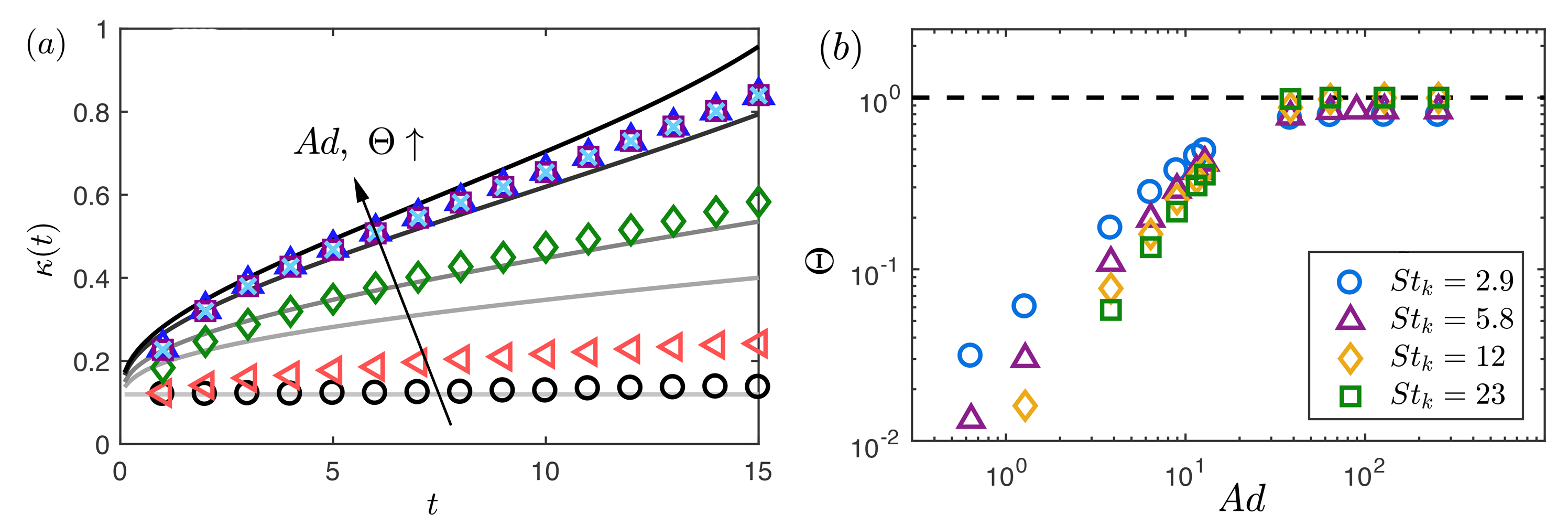}
  \centering
  \caption{\label{fig:ad} (a) Temporal evolution of the parameter $\kappa$ for DNS-DEM simulation with $St_k = 5.8$ and $Ad = 0.013$ (circles), $1.3$ (left-pointing triangles), $13$ (diamonds), $64$ (upward triangles), $128$ (squares), and $256$ (axes). The solid lines spanning from light to dark color are results from PBE with the sticking probability $\Theta =0$, $0.2$, $0.4$, $0.8$, and $1$. (b) Sticking probability $\Theta$, determined from Eq. (\ref{eq:theta_ad}), as a function of adhesion parameter $Ad$ for $St_k = 2.9$ (circles), $5.8$ (triangles), $12$ (diamonds), and $23$ (squares). The horizontal dashed line is $\Theta = 1$.}
\end{figure}

We solve PBE using agglomeration kernel $\Gamma_a(i,j)$ with $\Theta$ increasing from $0$ to $1$ (see Eq. (\ref{eq:sticking})). The evolution of the scale parameter $\kappa$ is shown as solid lines in Fig. \ref{fig:ad}(a). It is evident from the results that a smaller sticking probability $\Theta$ leads to a lower growth rate of agglomerates. We also plot corresponding results from DNS-DEM simulations with different values of adhesion parameter $Ad$ as data points in Fig. \ref{fig:ad}(a). For $Ad = 0.013$, $\kappa(t)$ is close to the PBE results with sticking probability $\Theta=0$, indicating that almost no agglomerates are formed given such a weak adhesive force. As $Ad$ increases beyond $\sim 64$, the $\kappa(t)$ curves converge to the PBE result with sticking probability $\Theta=1$. This strong adhesion case corresponds to the conventional PBE simulations, where the hit-and-stick assumption is made. Our results here suggest that PBE can also simulate the agglomeration process for particles with relatively weak adhesion once the sticking probability $\Theta$ is adopted.

We then determine the value of the sticking probability $\Theta$ in a statistical manner based on our DNS-DEM data. For a given $Ad$, we extract the instantaneous value of the scaling parameter $\kappa(t,Ad)$ from DNS-DEM simulations and map this point out on Fig. \ref{fig:ad}(a) and find the PBE curve of $\kappa(t,\Theta)$ that the point sits on. This procedure instantaneously correlates $\Theta(t)$ to $Ad$. Then time-averaging is performed to get the sticking probability at this given $Ad$:
\begin{equation}
  \label{eq:theta_ad}
   \Theta(Ad) = \frac{1}{T} \int_{0}^{T} \Theta(t) {\rm d}t.
\end{equation}
In Fig. \ref{fig:ad}(b), we plot $\Theta(Ad)$ for $St_k = 2.9$, $5.8$, $12$, and $23$. With $Ad < 1$, the sticking probability $\Theta$ for any $St_k$ is smaller than $\sim 0.3\%$ and the data points of different $St_k$ are rather scattered. In contrast, when $Ad > 10$, there is an adhesion-controlled regime, in which  $\Theta$ is mainly determined by $Ad$. Particularly, the unit sticking probability, $\Theta \approx 1$, which corresponds to the hit-and-stick situation, is achieved when $Ad$ is larger than $\sim 50$.

\subsection{Modelling sticking probability $\Theta$}
\label{sec_theta}
\begin{figure}
  \includegraphics[width = 9 cm]{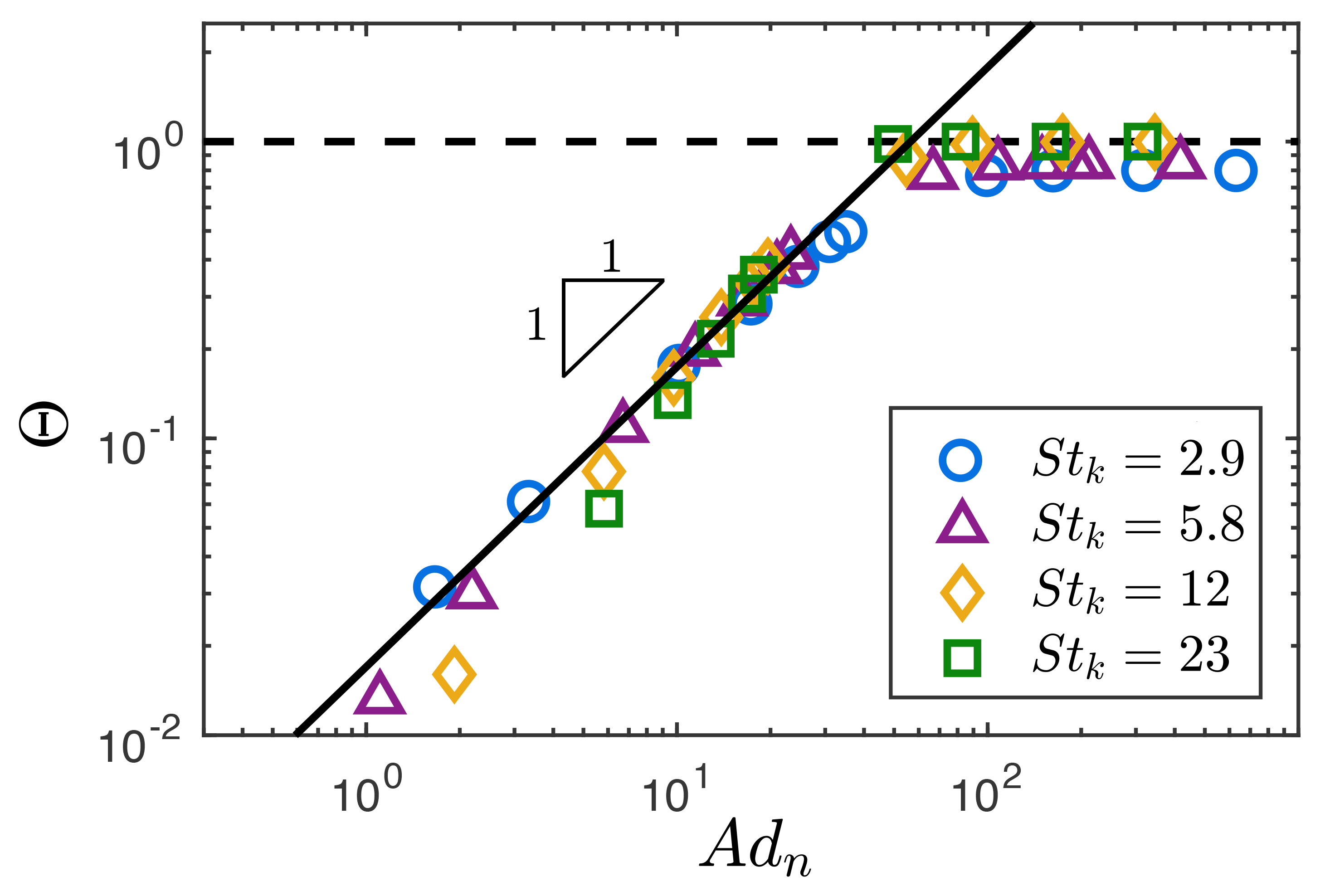}
  \centering
  \caption{\label{fig:adn} {\color{black}{Sticking probability $\Theta$ as a function of the adhesion parameter $Ad_n$, which is defined based on the averaged normal collision velocity $\langle v_{cn} \rangle$, for $St_k = 2.9$ (circles), $5.8$ (triangles), $12$ (diamonds), and $23$ (squares). The solid line is $\Theta = 0.017Ad_n$ and the horizontal dashed line is $\Theta = 1$.}}}
\end{figure}

{\color{black}{Describing turbulence-induced agglomeration using PBE requires knowledge of the sticking probability $\Theta$ {\it a priori}. Therefore, it is of significance to relate $\Theta$ to the particle-level properties. We consider a head-on collision between two primary particles with $v_{cn}$ being the relative collision velocity. For simplicity, only the normal forces in Eq. (\ref{eq:dem_a}) are taken into account and the interparticle overlap $\delta$ evolves according to
\begin{equation}
  \label{eqode}
\frac{\mathrm{d}^2\delta}{\mathrm{d}t^2} + \frac{2\eta_N}{m}\frac{\mathrm{d}\delta}{\mathrm{d}t} + \frac{8F_C}{m}\left( \hat{a}^3(\delta) - \hat{a}^{3/2} (\delta)\right) = 0,
\end{equation}
with the initial conditions $\delta(0) = 0$ and $\frac{\mathrm{d}\delta}{\mathrm{d}t}=v_{cn}$. The contact between the particles is built up when $\delta > 0$ and is broken when $\delta< -\delta_C$. Normalizing the overlap using its critical value $\delta_C$ and the time using $\delta_C/v_{cn}$, we have the following non-dimensional form of Eq. (\ref{eqode}):
\begin{equation}
  \label{eqode2}
  \frac{\mathrm{d}^2\hat{\delta}}{\mathrm{d}\hat{t}^2} + B \alpha \hat{a}^{1/2}\frac{\mathrm{d}\hat{\delta}}{\mathrm{d}\hat{t}} + 3.63B^2g(\hat{\delta}) = 0.
\end{equation}
The damping coefficient $\alpha$ is an input parameter and the scaled radius $\hat{a}$ can be calculated inversely through Eq. (\ref{eq:delta_a}). The results of a collision are determined by the parameter $B$, which is defined as
\begin{equation}
    \label{eq_B}
  B = 2.24\left( \frac{E_{ij}}{\rho_p v_{cn}^2} \right)^{-\frac{1}{3}}\left(\frac{\gamma}{\rho_p v_{cn}^2 r_p} \right)^{\frac{5}{6}}
\end{equation}
From Eq. (\ref{eq_B}) and a simple dimensional analysis, it is obvious that the effect of the adhesion (i.e., the surface energy $\gamma$) on the sticking probability $\theta$ is determined by the dimensionless adhesion parameter $Ad(v_{cn}) \equiv \gamma/(\rho_p v_{cn}^2 r_p)$, which is defined based on the normal collision velocity $v_{cn}$. We measure the value of $v_{cn}$ for every collision event in each simulation run and use the mean value $\langle v_{cn} \rangle$ as the typical velocity scale. A modified adhesion parameter then is given as
\begin{equation}
  \label{eq_adn}
  Ad_n = \frac{\gamma}{\rho_p \langle v_{cn} \rangle^2 r_p}.
\end{equation}
In Fig. \ref{fig:adn}, we replot the data of Fig. \ref{fig:ad}(b) in the $\Theta - Ad_n$ plane and all the data points collapse onto two curves:
\begin{equation}
  \label{eq_fittheta}
  \Theta = 0.017Ad_n,\ {\rm for}\ 1<Ad_n<30,\ {\rm and}\ \Theta = 1\ {\rm for}\ Ad_n > 50.
\end{equation}
The results in Fig. \ref{fig:adn} indicate that the mean relative collision velocity is an appropriate choice to scale the effect of adhesion and the sticking probability $\Theta$ can be well estimated once $Ad_n$ is known. Here, the data points for cases with $Ad_n<1$ are neglected, since the sticking probability is less than $10^{-2}$, which is too small to ensure good statistics. It should be noted that current values of $\langle v_{cn} \rangle$ are measured from DNS-DEM. To avoid computationally expensive DNS-DEM calculation, one can also adopt analytical expressions to estimate the value of $\langle v_{cn} \rangle$ (see \cite{pan2013turbulence, pan2010relative, ayala2008effects, rani2014stochastic}). }}

\section{Conclusions}
{\color{black}{In summary, for adhesive inertial particles suspended in turbulence, we measure both the collision rate, the structure and the size distribution of early-stage agglomerates with varying adhesion. We find that the collision rate is significantly reduced due to the adhesion-induced agglomeration. As the value of adhesion parameter $Ad$ increases, the system reaches a strong-adhesion limit, in which the sticking probability for colliding particles is unit and further increase of $Ad$ does not affect the dynamics of agglomeration. We also find that the size distribution of early-stage agglomerates follows an exponential equation $n(A)/n_0 = \beta(\kappa)\exp(-A/\kappa)$ regardless of the adhesion force magnitude. The transient dynamics of agglomeration at early-stage thus can be characterized using a single scale parameter $\kappa$. This finding may help to reduce the computing complexity of the population balance equation (PBE) to that of monodisperse systems since only one parameter $\kappa$ needs to be solved. The evolution of $\kappa$ then serves as an indicator for the quantitative comparison between DNS-DEM and PBE simulations. We show that, by introducing an agglomeration kernel constructed in terms of gyration radius of agglomerates and a sticking probability $\Theta$, PBE can well reproduce the results of DNS-DEM. A relationship between the sticking probability and particle properties is then proposed based on the scaling analysis of the equation for head-on collisions.

There are several interesting directions for future study. First, the current work focuses on the early-stage agglomeration, where the breakage and the rearrangement of agglomerates are not significant. It is unclear to what extend the framework developed here can be extended to situations with large agglomerates \cite{Dizaji2018JFM}. It requires one to construct kernel functions that contain information about breakage and restructuring \cite{Spicer1996, Pandya1983}. Moreover, we fix the value of Taylor-microscale Reynolds number $Re_{\lambda}$ in the current work. It is reported that the relative velocity and the collision rate for inertial particles increase strongly with increasing $Re_{\lambda}$ \cite{WangJFM2000,ZhouJFM2001, IrelandJFM2016}. However, a stronger clustering effect may suppress the agglomeration \cite{liu2018cluster}. A quantitative characterization of competing effects of the increasing collision rate and the decreasing sticking probability as $Re_{\lambda}$ increases would be of great interest.}}

\section*{Acknowledgements}
SQL acknowledges support from the National Fund for Distinguished Young Scholars of China (51725601) and National Key Research and Development Program of China (2016YFB0600602). The authors thank Prof. Q. Yao at Tsinghua, Prof. L. Mädler at the
University of Bremen and Dr. W. Liu at University of Surrey for useful suggestions.

%



\balance


\bibliography{draft_agglomeration}

\begin{thebibliography}{69}%
\makeatletter
\providecommand \@ifxundefined [1]{%
 \@ifx{#1\undefined}
}%
\providecommand \@ifnum [1]{%
 \ifnum #1\expandafter \@firstoftwo
 \else \expandafter \@secondoftwo
 \fi
}%
\providecommand \@ifx [1]{%
 \ifx #1\expandafter \@firstoftwo
 \else \expandafter \@secondoftwo
 \fi
}%
\providecommand \natexlab [1]{#1}%
\providecommand \enquote  [1]{``#1''}%
\providecommand \bibnamefont  [1]{#1}%
\providecommand \bibfnamefont [1]{#1}%
\providecommand \citenamefont [1]{#1}%
\providecommand \href@noop [0]{\@secondoftwo}%
\providecommand \href [0]{\begingroup \@sanitize@url \@href}%
\providecommand \@href[1]{\@@startlink{#1}\@@href}%
\providecommand \@@href[1]{\endgroup#1\@@endlink}%
\providecommand \@sanitize@url [0]{\catcode `\\12\catcode `\$12\catcode
  `\&12\catcode `\#12\catcode `\^12\catcode `\_12\catcode `\%12\relax}%
\providecommand \@@startlink[1]{}%
\providecommand \@@endlink[0]{}%
\providecommand \url  [0]{\begingroup\@sanitize@url \@url }%
\providecommand \@url [1]{\endgroup\@href {#1}{\urlprefix }}%
\providecommand \urlprefix  [0]{URL }%
\providecommand \Eprint [0]{\href }%
\providecommand \doibase [0]{http://dx.doi.org/}%
\providecommand \selectlanguage [0]{\@gobble}%
\providecommand \bibinfo  [0]{\@secondoftwo}%
\providecommand \bibfield  [0]{\@secondoftwo}%
\providecommand \translation [1]{[#1]}%
\providecommand \BibitemOpen [0]{}%
\providecommand \bibitemStop [0]{}%
\providecommand \bibitemNoStop [0]{.\EOS\space}%
\providecommand \EOS [0]{\spacefactor3000\relax}%
\providecommand \BibitemShut  [1]{\csname bibitem#1\endcsname}%
\let\auto@bib@innerbib\@empty
\bibitem [{\citenamefont {Saw}\ \emph {et~al.}(2008)\citenamefont {Saw},
  \citenamefont {Shaw}, \citenamefont {Ayyalasomayajula}, \citenamefont
  {Chuang},\ and\ \citenamefont {Gylfason}}]{SawPRL2008}%
  \BibitemOpen
  \bibfield  {author} {\bibinfo {author} {\bibfnamefont {E.~W.}\ \bibnamefont
  {Saw}}, \bibinfo {author} {\bibfnamefont {R.~A.}\ \bibnamefont {Shaw}},
  \bibinfo {author} {\bibfnamefont {S.}~\bibnamefont {Ayyalasomayajula}},
  \bibinfo {author} {\bibfnamefont {P.~Y.}\ \bibnamefont {Chuang}}, \ and\
  \bibinfo {author} {\bibfnamefont {A.}~\bibnamefont {Gylfason}},\ }\bibfield
  {title} {\enquote {\bibinfo {title} {Inertial clustering of particles in
  high-{R}eynolds-number turbulence},}\ }\href@noop {} {\bibfield  {journal}
  {\bibinfo  {journal} {Physical Review Letters}\ }\textbf {\bibinfo {volume}
  {100}},\ \bibinfo {pages} {214501} (\bibinfo {year} {2008})}\BibitemShut
  {NoStop}%
\bibitem [{\citenamefont {Lu}\ \emph {et~al.}(2010)\citenamefont {Lu},
  \citenamefont {Nordsiek}, \citenamefont {Saw},\ and\ \citenamefont
  {Shaw}}]{LuPRL2010}%
  \BibitemOpen
  \bibfield  {author} {\bibinfo {author} {\bibfnamefont {J.}~\bibnamefont
  {Lu}}, \bibinfo {author} {\bibfnamefont {H.}~\bibnamefont {Nordsiek}},
  \bibinfo {author} {\bibfnamefont {E.~W.}\ \bibnamefont {Saw}}, \ and\
  \bibinfo {author} {\bibfnamefont {R.~A.}\ \bibnamefont {Shaw}},\ }\bibfield
  {title} {\enquote {\bibinfo {title} {Clustering of charged inertial particles
  in turbulence},}\ }\href@noop {} {\bibfield  {journal} {\bibinfo  {journal}
  {Physical Review Letters}\ }\textbf {\bibinfo {volume} {104}},\ \bibinfo
  {pages} {184505} (\bibinfo {year} {2010})}\BibitemShut {NoStop}%
\bibitem [{\citenamefont {Bec}\ \emph {et~al.}(2014)\citenamefont {Bec},
  \citenamefont {Homann},\ and\ \citenamefont {Ray}}]{bec2014gravity}%
  \BibitemOpen
  \bibfield  {author} {\bibinfo {author} {\bibfnamefont {J.}~\bibnamefont
  {Bec}}, \bibinfo {author} {\bibfnamefont {H.}~\bibnamefont {Homann}}, \ and\
  \bibinfo {author} {\bibfnamefont {S.~S.}\ \bibnamefont {Ray}},\ }\bibfield
  {title} {\enquote {\bibinfo {title} {Gravity-driven enhancement of heavy
  particle clustering in turbulent flow},}\ }\href@noop {} {\bibfield
  {journal} {\bibinfo  {journal} {Physical Review Letters}\ }\textbf {\bibinfo
  {volume} {112}},\ \bibinfo {pages} {184501} (\bibinfo {year}
  {2014})}\BibitemShut {NoStop}%
\bibitem [{\citenamefont {Gustavsson}\ \emph {et~al.}(2014)\citenamefont
  {Gustavsson}, \citenamefont {Vajedi},\ and\ \citenamefont
  {Mehlig}}]{Mehlig2014}%
  \BibitemOpen
  \bibfield  {author} {\bibinfo {author} {\bibfnamefont {K.}~\bibnamefont
  {Gustavsson}}, \bibinfo {author} {\bibfnamefont {S.}~\bibnamefont {Vajedi}},
  \ and\ \bibinfo {author} {\bibfnamefont {B.}~\bibnamefont {Mehlig}},\
  }\bibfield  {title} {\enquote {\bibinfo {title} {Clustering of particles
  falling in a turbulent flow},}\ }\href@noop {} {\bibfield  {journal}
  {\bibinfo  {journal} {Physical Review Letters}\ }\textbf {\bibinfo {volume}
  {112}},\ \bibinfo {pages} {214501} (\bibinfo {year} {2014})}\BibitemShut
  {NoStop}%
\bibitem [{\citenamefont {Balachandar}\ and\ \citenamefont
  {Eaton}(2010)}]{balachandarAR2010}%
  \BibitemOpen
  \bibfield  {author} {\bibinfo {author} {\bibfnamefont {S.}~\bibnamefont
  {Balachandar}}\ and\ \bibinfo {author} {\bibfnamefont {J.~K.}\ \bibnamefont
  {Eaton}},\ }\bibfield  {title} {\enquote {\bibinfo {title} {Turbulent
  dispersed multiphase flow},}\ }\href@noop {} {\bibfield  {journal} {\bibinfo
  {journal} {Annual Review of Fluid Mechanics}\ }\textbf {\bibinfo {volume}
  {42}},\ \bibinfo {pages} {111--133} (\bibinfo {year} {2010})}\BibitemShut
  {NoStop}%
\bibitem [{\citenamefont {Smoluchowski}(1917)}]{Von1916}%
  \BibitemOpen
  \bibfield  {author} {\bibinfo {author} {\bibfnamefont {M}~\bibnamefont
  {Smoluchowski}},\ }\bibfield  {title} {\enquote {\bibinfo {title} {Versuch
  einer mathematischen theorie der koagulationskinetik kolloider
  l{\"o}sungen},}\ }\href@noop {} {\bibfield  {journal} {\bibinfo  {journal}
  {Zeitschrift f{\"u}r physikalische Chemie}\ }\textbf {\bibinfo {volume}
  {92}},\ \bibinfo {pages} {129--168} (\bibinfo {year} {1917})}\BibitemShut
  {NoStop}%
\bibitem [{\citenamefont {Friedlander}(2000)}]{friedlander2000}%
  \BibitemOpen
  \bibfield  {author} {\bibinfo {author} {\bibfnamefont {S.~K.}\ \bibnamefont
  {Friedlander}},\ }\href@noop {} {\emph {\bibinfo {title} {Smoke, Dust, and
  Haze: Fundamentals of Aerosol Dynamics}}}\ (\bibinfo {year}
  {2000})\BibitemShut {NoStop}%
\bibitem [{\citenamefont {Pumir}\ and\ \citenamefont
  {Wilkinson}(2016)}]{pumir2016}%
  \BibitemOpen
  \bibfield  {author} {\bibinfo {author} {\bibfnamefont {A.}~\bibnamefont
  {Pumir}}\ and\ \bibinfo {author} {\bibfnamefont {M.}~\bibnamefont
  {Wilkinson}},\ }\bibfield  {title} {\enquote {\bibinfo {title} {Collisional
  aggregation due to turbulence},}\ }\href@noop {} {\bibfield  {journal}
  {\bibinfo  {journal} {Annual Review of Condensed Matter Physics}\ }\textbf
  {\bibinfo {volume} {7}},\ \bibinfo {pages} {141--170} (\bibinfo {year}
  {2016})}\BibitemShut {NoStop}%
\bibitem [{\citenamefont {Saffman}\ and\ \citenamefont
  {Turner}(1956)}]{SaffmanJFM1956}%
  \BibitemOpen
  \bibfield  {author} {\bibinfo {author} {\bibfnamefont {P.G.F.}\ \bibnamefont
  {Saffman}}\ and\ \bibinfo {author} {\bibfnamefont {J.S.}\ \bibnamefont
  {Turner}},\ }\bibfield  {title} {\enquote {\bibinfo {title} {On the collision
  of drops in turbulent clouds},}\ }\href@noop {} {\bibfield  {journal}
  {\bibinfo  {journal} {Journal of Fluid Mechanics}\ }\textbf {\bibinfo
  {volume} {1}},\ \bibinfo {pages} {16--30} (\bibinfo {year}
  {1956})}\BibitemShut {NoStop}%
\bibitem [{\citenamefont {Abrahamson}(1975)}]{AbrahamsonCES1975}%
  \BibitemOpen
  \bibfield  {author} {\bibinfo {author} {\bibfnamefont {J.}~\bibnamefont
  {Abrahamson}},\ }\bibfield  {title} {\enquote {\bibinfo {title} {Collision
  rates of small particles in a vigorously turbulent fluid},}\ }\href@noop {}
  {\bibfield  {journal} {\bibinfo  {journal} {Chemical Engineering Science}\
  }\textbf {\bibinfo {volume} {30}},\ \bibinfo {pages} {1371--1379} (\bibinfo
  {year} {1975})}\BibitemShut {NoStop}%
\bibitem [{\citenamefont {Sundaram}\ and\ \citenamefont
  {Collins}(1997)}]{SundaramJFM1997}%
  \BibitemOpen
  \bibfield  {author} {\bibinfo {author} {\bibfnamefont {S.}~\bibnamefont
  {Sundaram}}\ and\ \bibinfo {author} {\bibfnamefont {L.~R.}\ \bibnamefont
  {Collins}},\ }\bibfield  {title} {\enquote {\bibinfo {title} {Collision
  statistics in an isotropic particle-laden turbulent suspension. {P}art 1.
  direct numerical simulations},}\ }\href@noop {} {\bibfield  {journal}
  {\bibinfo  {journal} {Journal of Fluid Mechanics}\ }\textbf {\bibinfo
  {volume} {335}},\ \bibinfo {pages} {75--109} (\bibinfo {year}
  {1997})}\BibitemShut {NoStop}%
\bibitem [{\citenamefont {Wang}\ \emph {et~al.}(2000)\citenamefont {Wang},
  \citenamefont {Wexler},\ and\ \citenamefont {Zhou}}]{WangJFM2000}%
  \BibitemOpen
  \bibfield  {author} {\bibinfo {author} {\bibfnamefont {L.~P.}\ \bibnamefont
  {Wang}}, \bibinfo {author} {\bibfnamefont {A.~S.}\ \bibnamefont {Wexler}}, \
  and\ \bibinfo {author} {\bibfnamefont {Y.}~\bibnamefont {Zhou}},\ }\bibfield
  {title} {\enquote {\bibinfo {title} {Statistical mechanical description and
  modelling of turbulent collision of inertial particles},}\ }\href@noop {}
  {\bibfield  {journal} {\bibinfo  {journal} {Journal of Fluid Mechanics}\
  }\textbf {\bibinfo {volume} {415}},\ \bibinfo {pages} {117--153} (\bibinfo
  {year} {2000})}\BibitemShut {NoStop}%
\bibitem [{\citenamefont {Zhou}\ \emph {et~al.}(2001)\citenamefont {Zhou},
  \citenamefont {Wexler},\ and\ \citenamefont {Wang}}]{ZhouJFM2001}%
  \BibitemOpen
  \bibfield  {author} {\bibinfo {author} {\bibfnamefont {Y.}~\bibnamefont
  {Zhou}}, \bibinfo {author} {\bibfnamefont {A.~S.}\ \bibnamefont {Wexler}}, \
  and\ \bibinfo {author} {\bibfnamefont {L.~P.}\ \bibnamefont {Wang}},\
  }\bibfield  {title} {\enquote {\bibinfo {title} {Modelling turbulent
  collision of bidisperse inertial particles},}\ }\href@noop {} {\bibfield
  {journal} {\bibinfo  {journal} {Journal of Fluid Mechanics}\ }\textbf
  {\bibinfo {volume} {433}},\ \bibinfo {pages} {77--104} (\bibinfo {year}
  {2001})}\BibitemShut {NoStop}%
\bibitem [{\citenamefont {Gustavsson}\ and\ \citenamefont
  {Mehlig}(2016)}]{gustavsson2016}%
  \BibitemOpen
  \bibfield  {author} {\bibinfo {author} {\bibfnamefont {K.}~\bibnamefont
  {Gustavsson}}\ and\ \bibinfo {author} {\bibfnamefont {B.}~\bibnamefont
  {Mehlig}},\ }\bibfield  {title} {\enquote {\bibinfo {title} {Statistical
  model for collisions and recollisions of inertial particles in mixing
  flows},}\ }\href@noop {} {\bibfield  {journal} {\bibinfo  {journal} {The
  European Physical Journal E}\ }\textbf {\bibinfo {volume} {39}},\ \bibinfo
  {pages} {55} (\bibinfo {year} {2016})}\BibitemShut {NoStop}%
\bibitem [{\citenamefont {Wilkinson}\ and\ \citenamefont
  {Mehlig}(2005)}]{wilkinson2005caustics}%
  \BibitemOpen
  \bibfield  {author} {\bibinfo {author} {\bibfnamefont {M.}~\bibnamefont
  {Wilkinson}}\ and\ \bibinfo {author} {\bibfnamefont {B.}~\bibnamefont
  {Mehlig}},\ }\bibfield  {title} {\enquote {\bibinfo {title} {Caustics in
  turbulent aerosols},}\ }\href@noop {} {\bibfield  {journal} {\bibinfo
  {journal} {Europhysics Letters}\ }\textbf {\bibinfo {volume} {71}},\ \bibinfo
  {pages} {186} (\bibinfo {year} {2005})}\BibitemShut {NoStop}%
\bibitem [{\citenamefont {Gustavsson}\ and\ \citenamefont
  {Mehlig}(2011)}]{gustavsson2011}%
  \BibitemOpen
  \bibfield  {author} {\bibinfo {author} {\bibfnamefont {K.}~\bibnamefont
  {Gustavsson}}\ and\ \bibinfo {author} {\bibfnamefont {B.}~\bibnamefont
  {Mehlig}},\ }\bibfield  {title} {\enquote {\bibinfo {title} {Distribution of
  relative velocities in turbulent aerosols},}\ }\href@noop {} {\bibfield
  {journal} {\bibinfo  {journal} {Physical Review E}\ }\textbf {\bibinfo
  {volume} {84}},\ \bibinfo {pages} {045304} (\bibinfo {year}
  {2011})}\BibitemShut {NoStop}%
\bibitem [{\citenamefont {Falkovich}\ and\ \citenamefont
  {Pumir}(2007)}]{falkovich2007}%
  \BibitemOpen
  \bibfield  {author} {\bibinfo {author} {\bibfnamefont {G.}~\bibnamefont
  {Falkovich}}\ and\ \bibinfo {author} {\bibfnamefont {A.}~\bibnamefont
  {Pumir}},\ }\bibfield  {title} {\enquote {\bibinfo {title} {Sling effect in
  collisions of water droplets in turbulent clouds},}\ }\href@noop {}
  {\bibfield  {journal} {\bibinfo  {journal} {Journal of the Atmospheric
  Sciences}\ }\textbf {\bibinfo {volume} {64}},\ \bibinfo {pages} {4497--4505}
  (\bibinfo {year} {2007})}\BibitemShut {NoStop}%
\bibitem [{\citenamefont {Jaworek}\ \emph {et~al.}(2018)\citenamefont
  {Jaworek}, \citenamefont {Marchewicz}, \citenamefont {Sobczyk}, \citenamefont
  {Krupa},\ and\ \citenamefont {Czech}}]{JaworekPECS2018}%
  \BibitemOpen
  \bibfield  {author} {\bibinfo {author} {\bibfnamefont {A.}~\bibnamefont
  {Jaworek}}, \bibinfo {author} {\bibfnamefont {A.}~\bibnamefont {Marchewicz}},
  \bibinfo {author} {\bibfnamefont {A.~T.}\ \bibnamefont {Sobczyk}}, \bibinfo
  {author} {\bibfnamefont {A.}~\bibnamefont {Krupa}}, \ and\ \bibinfo {author}
  {\bibfnamefont {T.}~\bibnamefont {Czech}},\ }\bibfield  {title} {\enquote
  {\bibinfo {title} {Two-stage electrostatic precipitators for the reduction of
  {PM}2.5 particle emission},}\ }\href@noop {} {\bibfield  {journal} {\bibinfo
  {journal} {Progress in Energy and Combustion Science}\ }\textbf {\bibinfo
  {volume} {67}},\ \bibinfo {pages} {206--233} (\bibinfo {year}
  {2018})}\BibitemShut {NoStop}%
\bibitem [{\citenamefont {Jarvis}\ \emph {et~al.}(2005)\citenamefont {Jarvis},
  \citenamefont {Jefferson}, \citenamefont {Gregory},\ and\ \citenamefont
  {Parsons}}]{JarvisWR2005}%
  \BibitemOpen
  \bibfield  {author} {\bibinfo {author} {\bibfnamefont {P.}~\bibnamefont
  {Jarvis}}, \bibinfo {author} {\bibfnamefont {B.}~\bibnamefont {Jefferson}},
  \bibinfo {author} {\bibfnamefont {J.}~\bibnamefont {Gregory}}, \ and\
  \bibinfo {author} {\bibfnamefont {S.~A.}\ \bibnamefont {Parsons}},\
  }\bibfield  {title} {\enquote {\bibinfo {title} {A review of floc strength
  and breakage},}\ }\href@noop {} {\bibfield  {journal} {\bibinfo  {journal}
  {Water Research}\ }\textbf {\bibinfo {volume} {39}},\ \bibinfo {pages}
  {3121--3137} (\bibinfo {year} {2005})}\BibitemShut {NoStop}%
\bibitem [{\citenamefont {Blum}\ \emph {et~al.}(2000)\citenamefont {Blum},
  \citenamefont {Wurm}, \citenamefont {Kempf}, \citenamefont {Poppe},
  \citenamefont {Klahr}, \citenamefont {Kozasa}, \citenamefont {Rott},
  \citenamefont {Henning}, \citenamefont {Dorschner}, \citenamefont
  {Schr{\"a}pler} \emph {et~al.}}]{blum2000}%
  \BibitemOpen
  \bibfield  {author} {\bibinfo {author} {\bibfnamefont {J.}~\bibnamefont
  {Blum}}, \bibinfo {author} {\bibfnamefont {G.}~\bibnamefont {Wurm}}, \bibinfo
  {author} {\bibfnamefont {S.}~\bibnamefont {Kempf}}, \bibinfo {author}
  {\bibfnamefont {T.}~\bibnamefont {Poppe}}, \bibinfo {author} {\bibfnamefont
  {H.}~\bibnamefont {Klahr}}, \bibinfo {author} {\bibfnamefont
  {T.}~\bibnamefont {Kozasa}}, \bibinfo {author} {\bibfnamefont
  {M.}~\bibnamefont {Rott}}, \bibinfo {author} {\bibfnamefont {T.}~\bibnamefont
  {Henning}}, \bibinfo {author} {\bibfnamefont {J.}~\bibnamefont {Dorschner}},
  \bibinfo {author} {\bibfnamefont {R.}~\bibnamefont {Schr{\"a}pler}},  \emph
  {et~al.},\ }\bibfield  {title} {\enquote {\bibinfo {title} {Growth and form
  of planetary seedlings: Results from a microgravity aggregation
  experiment},}\ }\href@noop {} {\bibfield  {journal} {\bibinfo  {journal}
  {Physical Review Letters}\ }\textbf {\bibinfo {volume} {85}},\ \bibinfo
  {pages} {2426} (\bibinfo {year} {2000})}\BibitemShut {NoStop}%
\bibitem [{\citenamefont {Tien}\ \emph {et~al.}(1977)\citenamefont {Tien},
  \citenamefont {Wang},\ and\ \citenamefont {Barot}}]{tien1977}%
  \BibitemOpen
  \bibfield  {author} {\bibinfo {author} {\bibfnamefont {C.}~\bibnamefont
  {Tien}}, \bibinfo {author} {\bibfnamefont {C.~S.}\ \bibnamefont {Wang}}, \
  and\ \bibinfo {author} {\bibfnamefont {D.~T.}\ \bibnamefont {Barot}},\
  }\bibfield  {title} {\enquote {\bibinfo {title} {Chainlike formation of
  particle deposits in fluid-particle separation},}\ }\href@noop {} {\bibfield
  {journal} {\bibinfo  {journal} {Science}\ }\textbf {\bibinfo {volume}
  {196}},\ \bibinfo {pages} {983--985} (\bibinfo {year} {1977})}\BibitemShut
  {NoStop}%
\bibitem [{\citenamefont {Chen}\ \emph {et~al.}(2016)\citenamefont {Chen},
  \citenamefont {Liu},\ and\ \citenamefont {Li}}]{ChenPRE2016}%
  \BibitemOpen
  \bibfield  {author} {\bibinfo {author} {\bibfnamefont {S.}~\bibnamefont
  {Chen}}, \bibinfo {author} {\bibfnamefont {W.}~\bibnamefont {Liu}}, \ and\
  \bibinfo {author} {\bibfnamefont {S.~Q.}\ \bibnamefont {Li}},\ }\bibfield
  {title} {\enquote {\bibinfo {title} {Effect of long-range electrostatic
  repulsion on pore clogging during microfiltration},}\ }\href@noop {}
  {\bibfield  {journal} {\bibinfo  {journal} {Physical Review E}\ }\textbf
  {\bibinfo {volume} {94}},\ \bibinfo {pages} {063108} (\bibinfo {year}
  {2016})}\BibitemShut {NoStop}%
\bibitem [{\citenamefont {Marshall}\ and\ \citenamefont
  {Li}(2014)}]{Marshall2014}%
  \BibitemOpen
  \bibfield  {author} {\bibinfo {author} {\bibfnamefont {J.~S.}\ \bibnamefont
  {Marshall}}\ and\ \bibinfo {author} {\bibfnamefont {S.~Q.}\ \bibnamefont
  {Li}},\ }\href@noop {} {\emph {\bibinfo {title} {Adhesive Particle Flow}}}\
  (\bibinfo  {publisher} {Cambridge University Press},\ \bibinfo {year}
  {2014})\BibitemShut {NoStop}%
\bibitem [{\citenamefont {Bec}\ \emph {et~al.}(2013)\citenamefont {Bec},
  \citenamefont {Musacchio},\ and\ \citenamefont {Ray}}]{bec2013sticky}%
  \BibitemOpen
  \bibfield  {author} {\bibinfo {author} {\bibfnamefont {J.}~\bibnamefont
  {Bec}}, \bibinfo {author} {\bibfnamefont {S.}~\bibnamefont {Musacchio}}, \
  and\ \bibinfo {author} {\bibfnamefont {S.~S.}\ \bibnamefont {Ray}},\
  }\bibfield  {title} {\enquote {\bibinfo {title} {Sticky elastic
  collisions},}\ }\href@noop {} {\bibfield  {journal} {\bibinfo  {journal}
  {Physical Review E}\ }\textbf {\bibinfo {volume} {87}},\ \bibinfo {pages}
  {063013} (\bibinfo {year} {2013})}\BibitemShut {NoStop}%
\bibitem [{\citenamefont {Marshall}(2009)}]{MarshallJCP2009}%
  \BibitemOpen
  \bibfield  {author} {\bibinfo {author} {\bibfnamefont {J.~S.}\ \bibnamefont
  {Marshall}},\ }\bibfield  {title} {\enquote {\bibinfo {title}
  {Discrete-element modeling of particulate aerosol flows},}\ }\href@noop {}
  {\bibfield  {journal} {\bibinfo  {journal} {Journal of Computational
  Physics}\ }\textbf {\bibinfo {volume} {228}},\ \bibinfo {pages} {1541--1561}
  (\bibinfo {year} {2009})}\BibitemShut {NoStop}%
\bibitem [{\citenamefont {Dizaji}\ and\ \citenamefont
  {Marshall}(2017)}]{DizajiPT2017}%
  \BibitemOpen
  \bibfield  {author} {\bibinfo {author} {\bibfnamefont {F.~F.}\ \bibnamefont
  {Dizaji}}\ and\ \bibinfo {author} {\bibfnamefont {J.~S.}\ \bibnamefont
  {Marshall}},\ }\bibfield  {title} {\enquote {\bibinfo {title} {On the
  significance of two-way coupling in simulation of turbulent particle
  agglomeration},}\ }\href@noop {} {\bibfield  {journal} {\bibinfo  {journal}
  {Powder Technology}\ }\textbf {\bibinfo {volume} {318}},\ \bibinfo {pages}
  {83--94} (\bibinfo {year} {2017})}\BibitemShut {NoStop}%
\bibitem [{\citenamefont {Sommerfeld}(2001)}]{SommerfeldIJMF2001}%
  \BibitemOpen
  \bibfield  {author} {\bibinfo {author} {\bibfnamefont {M.}~\bibnamefont
  {Sommerfeld}},\ }\bibfield  {title} {\enquote {\bibinfo {title} {Validation
  of a stochastic lagrangian modelling approach for inter-particle collisions
  in homogeneous isotropic turbulence},}\ }\href@noop {} {\bibfield  {journal}
  {\bibinfo  {journal} {International Journal of Multiphase Flow}\ }\textbf
  {\bibinfo {volume} {27}},\ \bibinfo {pages} {1829--1858} (\bibinfo {year}
  {2001})}\BibitemShut {NoStop}%
\bibitem [{\citenamefont {Almohammed}\ and\ \citenamefont
  {Breuer}(2016)}]{AlmohammedPT2016}%
  \BibitemOpen
  \bibfield  {author} {\bibinfo {author} {\bibfnamefont {N.}~\bibnamefont
  {Almohammed}}\ and\ \bibinfo {author} {\bibfnamefont {M.}~\bibnamefont
  {Breuer}},\ }\bibfield  {title} {\enquote {\bibinfo {title} {Modeling and
  simulation of agglomeration in turbulent particle-laden flows: A comparison
  between energy-based and momentum-based agglomeration models},}\ }\href@noop
  {} {\bibfield  {journal} {\bibinfo  {journal} {Powder Technology}\ }\textbf
  {\bibinfo {volume} {294}},\ \bibinfo {pages} {373--402} (\bibinfo {year}
  {2016})}\BibitemShut {NoStop}%
\bibitem [{\citenamefont {Lundgren}(2003)}]{Lundgren2003}%
  \BibitemOpen
  \bibfield  {author} {\bibinfo {author} {\bibfnamefont {T.~S.}\ \bibnamefont
  {Lundgren}},\ }\href@noop {} {\emph {\bibinfo {title} {Linearly forces
  isotropic turbulence}}},\ \bibinfo {type} {Tech. Rep.}\ (\bibinfo
  {institution} {MINNESOTA UNIV MINNEAPOLIS},\ \bibinfo {year}
  {2003})\BibitemShut {NoStop}%
\bibitem [{\citenamefont {Rosales}\ and\ \citenamefont
  {Meneveau}(2005)}]{RosalesPOF2005}%
  \BibitemOpen
  \bibfield  {author} {\bibinfo {author} {\bibfnamefont {C.}~\bibnamefont
  {Rosales}}\ and\ \bibinfo {author} {\bibfnamefont {C.}~\bibnamefont
  {Meneveau}},\ }\bibfield  {title} {\enquote {\bibinfo {title} {Linear forcing
  in numerical simulations of isotropic turbulence: Physical space
  implementations and convergence properties},}\ }\href@noop {} {\bibfield
  {journal} {\bibinfo  {journal} {Physics of Fluids}\ }\textbf {\bibinfo
  {volume} {17}},\ \bibinfo {pages} {095106} (\bibinfo {year}
  {2005})}\BibitemShut {NoStop}%
\bibitem [{\citenamefont {Di~Felice}(1994)}]{di1994}%
  \BibitemOpen
  \bibfield  {author} {\bibinfo {author} {\bibfnamefont {R.}~\bibnamefont
  {Di~Felice}},\ }\bibfield  {title} {\enquote {\bibinfo {title} {The voidage
  function for fluid-particle interaction systems},}\ }\href@noop {} {\bibfield
   {journal} {\bibinfo  {journal} {International Journal of Multiphase Flow}\
  }\textbf {\bibinfo {volume} {20}},\ \bibinfo {pages} {153--159} (\bibinfo
  {year} {1994})}\BibitemShut {NoStop}%
\bibitem [{\citenamefont {Saffman}(1965)}]{SaffmanJFM1965}%
  \BibitemOpen
  \bibfield  {author} {\bibinfo {author} {\bibfnamefont {P.~G.~T.}\
  \bibnamefont {Saffman}},\ }\bibfield  {title} {\enquote {\bibinfo {title}
  {The lift on a small sphere in a slow shear flow},}\ }\href@noop {}
  {\bibfield  {journal} {\bibinfo  {journal} {Journal of Fluid Mechanics}\
  }\textbf {\bibinfo {volume} {22}},\ \bibinfo {pages} {385--400} (\bibinfo
  {year} {1965})}\BibitemShut {NoStop}%
\bibitem [{\citenamefont {Rubinow}\ and\ \citenamefont
  {Keller}(1961)}]{RubinowJFM1961}%
  \BibitemOpen
  \bibfield  {author} {\bibinfo {author} {\bibfnamefont {S.~I.}\ \bibnamefont
  {Rubinow}}\ and\ \bibinfo {author} {\bibfnamefont {J.~B.}\ \bibnamefont
  {Keller}},\ }\bibfield  {title} {\enquote {\bibinfo {title} {The transverse
  force on a spinning sphere moving in a viscous fluid},}\ }\href@noop {}
  {\bibfield  {journal} {\bibinfo  {journal} {Journal of Fluid Mechanics}\
  }\textbf {\bibinfo {volume} {11}},\ \bibinfo {pages} {447--459} (\bibinfo
  {year} {1961})}\BibitemShut {NoStop}%
\bibitem [{\citenamefont {Chen}\ \emph {et~al.}(2019)\citenamefont {Chen},
  \citenamefont {Liu},\ and\ \citenamefont {Li}}]{chen2019fast}%
  \BibitemOpen
  \bibfield  {author} {\bibinfo {author} {\bibfnamefont {S.}~\bibnamefont
  {Chen}}, \bibinfo {author} {\bibfnamefont {Wenwei}\ \bibnamefont {Liu}}, \
  and\ \bibinfo {author} {\bibfnamefont {Shuiqing}\ \bibnamefont {Li}},\
  }\bibfield  {title} {\enquote {\bibinfo {title} {A fast adhesive discrete
  element method for random packings of fine particles},}\ }\href@noop {}
  {\bibfield  {journal} {\bibinfo  {journal} {Chemical Engineering Science}\
  }\textbf {\bibinfo {volume} {193}},\ \bibinfo {pages} {336--345} (\bibinfo
  {year} {2019})}\BibitemShut {NoStop}%
\bibitem [{\citenamefont {S{\"u}mer}\ and\ \citenamefont
  {Sitti}(2008)}]{SumerJAST2008}%
  \BibitemOpen
  \bibfield  {author} {\bibinfo {author} {\bibfnamefont {B.}~\bibnamefont
  {S{\"u}mer}}\ and\ \bibinfo {author} {\bibfnamefont {M.}~\bibnamefont
  {Sitti}},\ }\bibfield  {title} {\enquote {\bibinfo {title} {Rolling and
  spinning friction characterization of fine particles using lateral force
  microscopy based contact pushing},}\ }\href@noop {} {\bibfield  {journal}
  {\bibinfo  {journal} {Journal of Adhesion Science and Technology}\ }\textbf
  {\bibinfo {volume} {22}},\ \bibinfo {pages} {481--506} (\bibinfo {year}
  {2008})}\BibitemShut {NoStop}%
\bibitem [{\citenamefont {Yang}\ \emph {et~al.}(2013)\citenamefont {Yang},
  \citenamefont {Li},\ and\ \citenamefont {Yao}}]{YangPT2013}%
  \BibitemOpen
  \bibfield  {author} {\bibinfo {author} {\bibfnamefont {Mengmeng}\
  \bibnamefont {Yang}}, \bibinfo {author} {\bibfnamefont {Shuiqing}\
  \bibnamefont {Li}}, \ and\ \bibinfo {author} {\bibfnamefont {Qiang}\
  \bibnamefont {Yao}},\ }\bibfield  {title} {\enquote {\bibinfo {title}
  {Mechanistic studies of initial deposition of fine adhesive particles on a
  fiber using discrete-element methods},}\ }\href@noop {} {\bibfield  {journal}
  {\bibinfo  {journal} {Powder Technology}\ }\textbf {\bibinfo {volume}
  {248}},\ \bibinfo {pages} {44--53} (\bibinfo {year} {2013})}\BibitemShut
  {NoStop}%
\bibitem [{\citenamefont {Li}\ and\ \citenamefont
  {Marshall}(2007)}]{LiJAS2007}%
  \BibitemOpen
  \bibfield  {author} {\bibinfo {author} {\bibfnamefont {S.~Q.}\ \bibnamefont
  {Li}}\ and\ \bibinfo {author} {\bibfnamefont {J.~S.}\ \bibnamefont
  {Marshall}},\ }\bibfield  {title} {\enquote {\bibinfo {title} {Discrete
  element simulation of micro-particle deposition on a cylindrical fiber in an
  array},}\ }\href@noop {} {\bibfield  {journal} {\bibinfo  {journal} {Journal
  of Aerosol Science}\ }\textbf {\bibinfo {volume} {38}},\ \bibinfo {pages}
  {1031--1046} (\bibinfo {year} {2007})}\BibitemShut {NoStop}%
\bibitem [{\citenamefont {Chen}\ \emph {et~al.}(2015)\citenamefont {Chen},
  \citenamefont {Li},\ and\ \citenamefont {Yang}}]{ChenPT2015}%
  \BibitemOpen
  \bibfield  {author} {\bibinfo {author} {\bibfnamefont {S.}~\bibnamefont
  {Chen}}, \bibinfo {author} {\bibfnamefont {S.~Q.}\ \bibnamefont {Li}}, \ and\
  \bibinfo {author} {\bibfnamefont {M.}~\bibnamefont {Yang}},\ }\bibfield
  {title} {\enquote {\bibinfo {title} {Sticking/rebound criterion for
  collisions of small adhesive particles: Effects of impact parameter and
  particle size},}\ }\href@noop {} {\bibfield  {journal} {\bibinfo  {journal}
  {Powder Technology}\ }\textbf {\bibinfo {volume} {274}},\ \bibinfo {pages}
  {431--440} (\bibinfo {year} {2015})}\BibitemShut {NoStop}%
\bibitem [{\citenamefont {Liu}\ \emph {et~al.}(2015)\citenamefont {Liu},
  \citenamefont {Li}, \citenamefont {Baule},\ and\ \citenamefont
  {Makse}}]{LiuSM2015}%
  \BibitemOpen
  \bibfield  {author} {\bibinfo {author} {\bibfnamefont {W.}~\bibnamefont
  {Liu}}, \bibinfo {author} {\bibfnamefont {S.~Q.}\ \bibnamefont {Li}},
  \bibinfo {author} {\bibfnamefont {A.}~\bibnamefont {Baule}}, \ and\ \bibinfo
  {author} {\bibfnamefont {H.~A.}\ \bibnamefont {Makse}},\ }\bibfield  {title}
  {\enquote {\bibinfo {title} {Adhesive loose packings of small dry
  particles},}\ }\href@noop {} {\bibfield  {journal} {\bibinfo  {journal} {Soft
  Matter}\ }\textbf {\bibinfo {volume} {11}},\ \bibinfo {pages} {6492--6498}
  (\bibinfo {year} {2015})}\BibitemShut {NoStop}%
\bibitem [{\citenamefont {Liu}\ \emph {et~al.}(2017)\citenamefont {Liu},
  \citenamefont {Jin}, \citenamefont {Chen}, \citenamefont {Makse},\ and\
  \citenamefont {Li}}]{LiuSM2017}%
  \BibitemOpen
  \bibfield  {author} {\bibinfo {author} {\bibfnamefont {W.}~\bibnamefont
  {Liu}}, \bibinfo {author} {\bibfnamefont {Y.}~\bibnamefont {Jin}}, \bibinfo
  {author} {\bibfnamefont {S.}~\bibnamefont {Chen}}, \bibinfo {author}
  {\bibfnamefont {H.~A.}\ \bibnamefont {Makse}}, \ and\ \bibinfo {author}
  {\bibfnamefont {S.~Q.}\ \bibnamefont {Li}},\ }\bibfield  {title} {\enquote
  {\bibinfo {title} {Equation of state for random sphere packings with
  arbitrary adhesion and friction},}\ }\href@noop {} {\bibfield  {journal}
  {\bibinfo  {journal} {Soft Matter}\ }\textbf {\bibinfo {volume} {13}},\
  \bibinfo {pages} {421--427} (\bibinfo {year} {2017})}\BibitemShut {NoStop}%
\bibitem [{\citenamefont {Marshall}(2011)}]{Marshall2011}%
  \BibitemOpen
  \bibfield  {author} {\bibinfo {author} {\bibfnamefont {J.~S.}\ \bibnamefont
  {Marshall}},\ }\bibfield  {title} {\enquote {\bibinfo {title} {Viscous
  damping force during head-on collision of two spherical particles},}\
  }\href@noop {} {\bibfield  {journal} {\bibinfo  {journal} {Physics of
  Fluids}\ }\textbf {\bibinfo {volume} {23}},\ \bibinfo {pages} {013305}
  (\bibinfo {year} {2011})}\BibitemShut {NoStop}%
\bibitem [{\citenamefont {Yang}\ and\ \citenamefont
  {Hunt}(2006)}]{YangPOF2006}%
  \BibitemOpen
  \bibfield  {author} {\bibinfo {author} {\bibfnamefont {F.~L.}\ \bibnamefont
  {Yang}}\ and\ \bibinfo {author} {\bibfnamefont {M.~L.}\ \bibnamefont
  {Hunt}},\ }\bibfield  {title} {\enquote {\bibinfo {title} {Dynamics of
  particle-particle collisions in a viscous liquid},}\ }\href@noop {}
  {\bibfield  {journal} {\bibinfo  {journal} {Physics of Fluids}\ }\textbf
  {\bibinfo {volume} {18}},\ \bibinfo {pages} {121506} (\bibinfo {year}
  {2006})}\BibitemShut {NoStop}%
\bibitem [{\citenamefont {Fayed}\ and\ \citenamefont
  {Ragab}(2013)}]{fayed2013}%
  \BibitemOpen
  \bibfield  {author} {\bibinfo {author} {\bibfnamefont {H.~E.}\ \bibnamefont
  {Fayed}}\ and\ \bibinfo {author} {\bibfnamefont {S.~A.}\ \bibnamefont
  {Ragab}},\ }\bibfield  {title} {\enquote {\bibinfo {title} {Direct numerical
  simulation of particles-bubbles collisions kernel in homogeneous isotropic
  turbulence},}\ }\href@noop {} {\bibfield  {journal} {\bibinfo  {journal} {The
  Journal of Computational Multiphase Flows}\ }\textbf {\bibinfo {volume}
  {5}},\ \bibinfo {pages} {167--188} (\bibinfo {year} {2013})}\BibitemShut
  {NoStop}%
\bibitem [{\citenamefont {Jin}\ and\ \citenamefont
  {Marshall}(2017)}]{JinJOE2017}%
  \BibitemOpen
  \bibfield  {author} {\bibinfo {author} {\bibfnamefont {X.}~\bibnamefont
  {Jin}}\ and\ \bibinfo {author} {\bibfnamefont {J.~S.}\ \bibnamefont
  {Marshall}},\ }\bibfield  {title} {\enquote {\bibinfo {title} {The role of
  fluid turbulence on contact electrification of suspended particles},}\
  }\href@noop {} {\bibfield  {journal} {\bibinfo  {journal} {Journal of
  Electrostatics}\ }\textbf {\bibinfo {volume} {87}},\ \bibinfo {pages}
  {217--227} (\bibinfo {year} {2017})}\BibitemShut {NoStop}%
\bibitem [{\citenamefont {Pruppacher}\ and\ \citenamefont
  {Klett}(1997)}]{pruppacher2012microphysics}%
  \BibitemOpen
  \bibfield  {author} {\bibinfo {author} {\bibfnamefont {Hans~R}\ \bibnamefont
  {Pruppacher}}\ and\ \bibinfo {author} {\bibfnamefont {James~D}\ \bibnamefont
  {Klett}},\ }\href@noop {} {\emph {\bibinfo {title} {Microphysics of Clouds
  and Precipitation}}}\ (\bibinfo  {publisher} {Kluwer},\ \bibinfo {year}
  {1997})\BibitemShut {NoStop}%
\bibitem [{\citenamefont {Onishi}\ \emph {et~al.}(2009)\citenamefont {Onishi},
  \citenamefont {Takahashi},\ and\ \citenamefont
  {Komori}}]{onishi2009influence}%
  \BibitemOpen
  \bibfield  {author} {\bibinfo {author} {\bibfnamefont {Ryo}\ \bibnamefont
  {Onishi}}, \bibinfo {author} {\bibfnamefont {Keiko}\ \bibnamefont
  {Takahashi}}, \ and\ \bibinfo {author} {\bibfnamefont {Satoru}\ \bibnamefont
  {Komori}},\ }\bibfield  {title} {\enquote {\bibinfo {title} {Influence of
  gravity on collisions of monodispersed droplets in homogeneous isotropic
  turbulence},}\ }\href@noop {} {\bibfield  {journal} {\bibinfo  {journal}
  {Physics of Fluids}\ }\textbf {\bibinfo {volume} {21}},\ \bibinfo {pages}
  {125108} (\bibinfo {year} {2009})}\BibitemShut {NoStop}%
\bibitem [{\citenamefont {Ireland}\ \emph
  {et~al.}(2016{\natexlab{a}})\citenamefont {Ireland}, \citenamefont {Bragg},\
  and\ \citenamefont {Collins}}]{ireland2016gravity}%
  \BibitemOpen
  \bibfield  {author} {\bibinfo {author} {\bibfnamefont {Peter~J}\ \bibnamefont
  {Ireland}}, \bibinfo {author} {\bibfnamefont {Andrew~D}\ \bibnamefont
  {Bragg}}, \ and\ \bibinfo {author} {\bibfnamefont {Lance~R}\ \bibnamefont
  {Collins}},\ }\bibfield  {title} {\enquote {\bibinfo {title} {The effect of
  {R}eynolds number on inertial particle dynamics in isotropic turbulence.
  {P}art 2. simulations with gravitational effects},}\ }\href@noop {}
  {\bibfield  {journal} {\bibinfo  {journal} {Journal of Fluid Mechanics}\
  }\textbf {\bibinfo {volume} {796}},\ \bibinfo {pages} {659--711} (\bibinfo
  {year} {2016}{\natexlab{a}})}\BibitemShut {NoStop}%
\bibitem [{\citenamefont {Krijt}\ \emph {et~al.}(2013)\citenamefont {Krijt},
  \citenamefont {G{\"u}ttler}, \citenamefont {Hei{\ss}elmann}, \citenamefont
  {Dominik},\ and\ \citenamefont {Tielens}}]{krijt2013energy}%
  \BibitemOpen
  \bibfield  {author} {\bibinfo {author} {\bibfnamefont {S.}~\bibnamefont
  {Krijt}}, \bibinfo {author} {\bibfnamefont {C.}~\bibnamefont {G{\"u}ttler}},
  \bibinfo {author} {\bibfnamefont {D.}~\bibnamefont {Hei{\ss}elmann}},
  \bibinfo {author} {\bibfnamefont {C.}~\bibnamefont {Dominik}}, \ and\
  \bibinfo {author} {\bibfnamefont {A.~G. G.~M.}\ \bibnamefont {Tielens}},\
  }\bibfield  {title} {\enquote {\bibinfo {title} {Energy dissipation in
  head-on collisions of spheres},}\ }\href@noop {} {\bibfield  {journal}
  {\bibinfo  {journal} {Journal of Physics D: Applied Physics}\ }\textbf
  {\bibinfo {volume} {46}},\ \bibinfo {pages} {435303} (\bibinfo {year}
  {2013})}\BibitemShut {NoStop}%
\bibitem [{\citenamefont {Chen}\ \emph {et~al.}(2018)\citenamefont {Chen},
  \citenamefont {Yau},\ and\ \citenamefont {Bartello}}]{chen2018turbulence}%
  \BibitemOpen
  \bibfield  {author} {\bibinfo {author} {\bibfnamefont {S.}~\bibnamefont
  {Chen}}, \bibinfo {author} {\bibfnamefont {M.~K.}\ \bibnamefont {Yau}}, \
  and\ \bibinfo {author} {\bibfnamefont {P.}~\bibnamefont {Bartello}},\
  }\bibfield  {title} {\enquote {\bibinfo {title} {Turbulence effects of
  collision efficiency and broadening of droplet size distribution in cumulus
  clouds},}\ }\href@noop {} {\bibfield  {journal} {\bibinfo  {journal} {Journal
  of the Atmospheric Sciences}\ }\textbf {\bibinfo {volume} {75}},\ \bibinfo
  {pages} {203--217} (\bibinfo {year} {2018})}\BibitemShut {NoStop}%
\bibitem [{\citenamefont {Dizaji}\ and\ \citenamefont
  {Marshall}(2016)}]{dizaji2016accelerated}%
  \BibitemOpen
  \bibfield  {author} {\bibinfo {author} {\bibfnamefont {F.~F.}\ \bibnamefont
  {Dizaji}}\ and\ \bibinfo {author} {\bibfnamefont {J.~S.}\ \bibnamefont
  {Marshall}},\ }\bibfield  {title} {\enquote {\bibinfo {title} {An accelerated
  stochastic vortex structure method for particle collision and agglomeration
  in homogeneous turbulence},}\ }\href@noop {} {\bibfield  {journal} {\bibinfo
  {journal} {Physics of Fluids}\ }\textbf {\bibinfo {volume} {28}},\ \bibinfo
  {pages} {113301} (\bibinfo {year} {2016})}\BibitemShut {NoStop}%
\bibitem [{\citenamefont {Jiang}\ and\ \citenamefont
  {Logan}(1991)}]{JiangEST1991}%
  \BibitemOpen
  \bibfield  {author} {\bibinfo {author} {\bibfnamefont {Q.}~\bibnamefont
  {Jiang}}\ and\ \bibinfo {author} {\bibfnamefont {B.~E.}\ \bibnamefont
  {Logan}},\ }\bibfield  {title} {\enquote {\bibinfo {title} {Fractal
  dimensions of aggregates determined from steady-state size distributions},}\
  }\href@noop {} {\bibfield  {journal} {\bibinfo  {journal} {Environmental
  Science \& Technology}\ }\textbf {\bibinfo {volume} {25}},\ \bibinfo {pages}
  {2031--2038} (\bibinfo {year} {1991})}\BibitemShut {NoStop}%
\bibitem [{\citenamefont {Flesch}\ \emph {et~al.}(1999)\citenamefont {Flesch},
  \citenamefont {Spicer},\ and\ \citenamefont {Pratsinis}}]{flesch1999laminar}%
  \BibitemOpen
  \bibfield  {author} {\bibinfo {author} {\bibfnamefont {J.~C.}\ \bibnamefont
  {Flesch}}, \bibinfo {author} {\bibfnamefont {P.~T.}\ \bibnamefont {Spicer}},
  \ and\ \bibinfo {author} {\bibfnamefont {S.~E.}\ \bibnamefont {Pratsinis}},\
  }\bibfield  {title} {\enquote {\bibinfo {title} {Laminar and turbulent
  shear-induced flocculation of fractal aggregates},}\ }\href@noop {}
  {\bibfield  {journal} {\bibinfo  {journal} {AIChE journal}\ }\textbf
  {\bibinfo {volume} {45}},\ \bibinfo {pages} {1114--1124} (\bibinfo {year}
  {1999})}\BibitemShut {NoStop}%
\bibitem [{\citenamefont {Waldner}\ \emph {et~al.}(2005)\citenamefont
  {Waldner}, \citenamefont {Sefcik}, \citenamefont {Soos},\ and\ \citenamefont
  {Morbidelli}}]{WaldnerPT2005}%
  \BibitemOpen
  \bibfield  {author} {\bibinfo {author} {\bibfnamefont {M.~H.}\ \bibnamefont
  {Waldner}}, \bibinfo {author} {\bibfnamefont {J.}~\bibnamefont {Sefcik}},
  \bibinfo {author} {\bibfnamefont {M.}~\bibnamefont {Soos}}, \ and\ \bibinfo
  {author} {\bibfnamefont {M.}~\bibnamefont {Morbidelli}},\ }\bibfield  {title}
  {\enquote {\bibinfo {title} {Initial growth kinetics and structure of
  colloidal aggregates in a turbulent coagulator},}\ }\href@noop {} {\bibfield
  {journal} {\bibinfo  {journal} {Powder Technology}\ }\textbf {\bibinfo
  {volume} {156}},\ \bibinfo {pages} {226--234} (\bibinfo {year}
  {2005})}\BibitemShut {NoStop}%
\bibitem [{\citenamefont {Selomulya}\ \emph {et~al.}(2001)\citenamefont
  {Selomulya}, \citenamefont {Amal}, \citenamefont {Bushell},\ and\
  \citenamefont {Waite}}]{selomulya2001evidence}%
  \BibitemOpen
  \bibfield  {author} {\bibinfo {author} {\bibfnamefont {Cordelia}\
  \bibnamefont {Selomulya}}, \bibinfo {author} {\bibfnamefont {Rose}\
  \bibnamefont {Amal}}, \bibinfo {author} {\bibfnamefont {Graeme}\ \bibnamefont
  {Bushell}}, \ and\ \bibinfo {author} {\bibfnamefont {T~David}\ \bibnamefont
  {Waite}},\ }\bibfield  {title} {\enquote {\bibinfo {title} {Evidence of shear
  rate dependence on restructuring and breakup of latex aggregates},}\
  }\href@noop {} {\bibfield  {journal} {\bibinfo  {journal} {Journal of Colloid
  and Interface Science}\ }\textbf {\bibinfo {volume} {236}},\ \bibinfo {pages}
  {67--77} (\bibinfo {year} {2001})}\BibitemShut {NoStop}%
\bibitem [{\citenamefont {Liu}\ \emph {et~al.}(2018)\citenamefont {Liu},
  \citenamefont {Wang}, \citenamefont {Chen},\ and\ \citenamefont
  {Liu}}]{LiuPT2018}%
  \BibitemOpen
  \bibfield  {author} {\bibinfo {author} {\bibfnamefont {D.}~\bibnamefont
  {Liu}}, \bibinfo {author} {\bibfnamefont {Z.}~\bibnamefont {Wang}}, \bibinfo
  {author} {\bibfnamefont {X.}~\bibnamefont {Chen}}, \ and\ \bibinfo {author}
  {\bibfnamefont {M.}~\bibnamefont {Liu}},\ }\bibfield  {title} {\enquote
  {\bibinfo {title} {Simulation of agglomerate breakage and restructuring in
  shear flows: Coupled effects of shear gradient, surface energy and initial
  structure},}\ }\href@noop {} {\bibfield  {journal} {\bibinfo  {journal}
  {Powder Technology}\ } (\bibinfo {year} {2018})}\BibitemShut {NoStop}%
\bibitem [{\citenamefont {Friedlander}\ and\ \citenamefont
  {Wang}(1966)}]{Friedlander1966}%
  \BibitemOpen
  \bibfield  {author} {\bibinfo {author} {\bibfnamefont {S.~K.}\ \bibnamefont
  {Friedlander}}\ and\ \bibinfo {author} {\bibfnamefont {C.~S.}\ \bibnamefont
  {Wang}},\ }\bibfield  {title} {\enquote {\bibinfo {title} {The
  self-preserving particle size distribution for coagulation by brownian
  motion},}\ }\href@noop {} {\bibfield  {journal} {\bibinfo  {journal} {Journal
  of Colloid and interface Science}\ }\textbf {\bibinfo {volume} {22}},\
  \bibinfo {pages} {126--132} (\bibinfo {year} {1966})}\BibitemShut {NoStop}%
\bibitem [{\citenamefont {Vemury}\ and\ \citenamefont
  {Pratsinis}(1995)}]{Vemury1995}%
  \BibitemOpen
  \bibfield  {author} {\bibinfo {author} {\bibfnamefont {S.}~\bibnamefont
  {Vemury}}\ and\ \bibinfo {author} {\bibfnamefont {S.~E.}\ \bibnamefont
  {Pratsinis}},\ }\bibfield  {title} {\enquote {\bibinfo {title}
  {Self-preserving size distributions of agglomerates},}\ }\href@noop {}
  {\bibfield  {journal} {\bibinfo  {journal} {Journal of Aerosol Science}\
  }\textbf {\bibinfo {volume} {26}},\ \bibinfo {pages} {175--185} (\bibinfo
  {year} {1995})}\BibitemShut {NoStop}%
\bibitem [{\citenamefont {Eggersdorfer}\ and\ \citenamefont
  {Pratsinis}(2014)}]{eggersdorfer2014}%
  \BibitemOpen
  \bibfield  {author} {\bibinfo {author} {\bibfnamefont {Maximilian~L}\
  \bibnamefont {Eggersdorfer}}\ and\ \bibinfo {author} {\bibfnamefont
  {Sotiris~E}\ \bibnamefont {Pratsinis}},\ }\bibfield  {title} {\enquote
  {\bibinfo {title} {Agglomerates and aggregates of nanoparticles made in the
  gas phase},}\ }\href@noop {} {\bibfield  {journal} {\bibinfo  {journal}
  {Advanced Powder Technology}\ }\textbf {\bibinfo {volume} {25}},\ \bibinfo
  {pages} {71--90} (\bibinfo {year} {2014})}\BibitemShut {NoStop}%
\bibitem [{\citenamefont {Jullien}\ and\ \citenamefont
  {Meakin}(1989)}]{Jullien1989}%
  \BibitemOpen
  \bibfield  {author} {\bibinfo {author} {\bibfnamefont {R.}~\bibnamefont
  {Jullien}}\ and\ \bibinfo {author} {\bibfnamefont {P.}~\bibnamefont
  {Meakin}},\ }\bibfield  {title} {\enquote {\bibinfo {title} {Simple models
  for the restructuring of three-dimensional ballistic aggregates},}\
  }\href@noop {} {\bibfield  {journal} {\bibinfo  {journal} {Journal of Colloid
  and Interface Science}\ }\textbf {\bibinfo {volume} {127}},\ \bibinfo {pages}
  {265--272} (\bibinfo {year} {1989})}\BibitemShut {NoStop}%
\bibitem [{\citenamefont {Elimelech}\ \emph {et~al.}(1998)\citenamefont
  {Elimelech}, \citenamefont {Jia}, \citenamefont {Gregory},\ and\
  \citenamefont {Williams}}]{Eimelech1998}%
  \BibitemOpen
  \bibfield  {author} {\bibinfo {author} {\bibfnamefont {M.}~\bibnamefont
  {Elimelech}}, \bibinfo {author} {\bibfnamefont {X.}~\bibnamefont {Jia}},
  \bibinfo {author} {\bibfnamefont {J.}~\bibnamefont {Gregory}}, \ and\
  \bibinfo {author} {\bibfnamefont {R.}~\bibnamefont {Williams}},\ }\href@noop
  {} {\emph {\bibinfo {title} {Particle Deposition and Aggregation:
  Measurement, Modelling and Simulation}}}\ (\bibinfo  {publisher} {Elsevier},\
  \bibinfo {year} {1998})\BibitemShut {NoStop}%
\bibitem [{\citenamefont {Pan}\ and\ \citenamefont
  {Padoan}(2013)}]{pan2013turbulence}%
  \BibitemOpen
  \bibfield  {author} {\bibinfo {author} {\bibfnamefont {Liubin}\ \bibnamefont
  {Pan}}\ and\ \bibinfo {author} {\bibfnamefont {Paolo}\ \bibnamefont
  {Padoan}},\ }\bibfield  {title} {\enquote {\bibinfo {title}
  {Turbulence-induced relative velocity of dust particles. i. identical
  particles},}\ }\href@noop {} {\bibfield  {journal} {\bibinfo  {journal} {The
  Astrophysical Journal}\ }\textbf {\bibinfo {volume} {776}},\ \bibinfo {pages}
  {12} (\bibinfo {year} {2013})}\BibitemShut {NoStop}%
\bibitem [{\citenamefont {Pan}\ and\ \citenamefont
  {Padoan}(2010)}]{pan2010relative}%
  \BibitemOpen
  \bibfield  {author} {\bibinfo {author} {\bibfnamefont {Liubin}\ \bibnamefont
  {Pan}}\ and\ \bibinfo {author} {\bibfnamefont {Paolo}\ \bibnamefont
  {Padoan}},\ }\bibfield  {title} {\enquote {\bibinfo {title} {Relative
  velocity of inertial particles in turbulent flows},}\ }\href@noop {}
  {\bibfield  {journal} {\bibinfo  {journal} {Journal of Fluid Mechanics}\
  }\textbf {\bibinfo {volume} {661}},\ \bibinfo {pages} {73--107} (\bibinfo
  {year} {2010})}\BibitemShut {NoStop}%
\bibitem [{\citenamefont {Ayala}\ \emph {et~al.}(2008)\citenamefont {Ayala},
  \citenamefont {Rosa},\ and\ \citenamefont {Wang}}]{ayala2008effects}%
  \BibitemOpen
  \bibfield  {author} {\bibinfo {author} {\bibfnamefont {Orlando}\ \bibnamefont
  {Ayala}}, \bibinfo {author} {\bibfnamefont {Bogdan}\ \bibnamefont {Rosa}}, \
  and\ \bibinfo {author} {\bibfnamefont {Lian-Ping}\ \bibnamefont {Wang}},\
  }\bibfield  {title} {\enquote {\bibinfo {title} {Effects of turbulence on the
  geometric collision rate of sedimenting droplets. {P}art 2. theory and
  parameterization},}\ }\href@noop {} {\bibfield  {journal} {\bibinfo
  {journal} {New Journal of Physics}\ }\textbf {\bibinfo {volume} {10}},\
  \bibinfo {pages} {075016} (\bibinfo {year} {2008})}\BibitemShut {NoStop}%
\bibitem [{\citenamefont {Rani}\ \emph {et~al.}(2014)\citenamefont {Rani},
  \citenamefont {Dhariwal},\ and\ \citenamefont {Koch}}]{rani2014stochastic}%
  \BibitemOpen
  \bibfield  {author} {\bibinfo {author} {\bibfnamefont {S.~L.}\ \bibnamefont
  {Rani}}, \bibinfo {author} {\bibfnamefont {R.}~\bibnamefont {Dhariwal}}, \
  and\ \bibinfo {author} {\bibfnamefont {D.~L.}\ \bibnamefont {Koch}},\
  }\bibfield  {title} {\enquote {\bibinfo {title} {A stochastic model for the
  relative motion of high stokes number particles in isotropic turbulence},}\
  }\href@noop {} {\bibfield  {journal} {\bibinfo  {journal} {Journal of Fluid
  Mechanics}\ }\textbf {\bibinfo {volume} {756}},\ \bibinfo {pages} {870--902}
  (\bibinfo {year} {2014})}\BibitemShut {NoStop}%
\bibitem [{\citenamefont {Dizaji}\ \emph {et~al.}(in press, 2019)\citenamefont
  {Dizaji}, \citenamefont {Marshall},\ and\ \citenamefont
  {Grant}}]{Dizaji2018JFM}%
  \BibitemOpen
  \bibfield  {author} {\bibinfo {author} {\bibfnamefont {F.}~\bibnamefont
  {Dizaji}}, \bibinfo {author} {\bibfnamefont {J.~S.}\ \bibnamefont
  {Marshall}}, \ and\ \bibinfo {author} {\bibfnamefont {J.~R.}\ \bibnamefont
  {Grant}},\ }\bibfield  {title} {\enquote {\bibinfo {title} {Collision and
  breakup of fractal particle agglomerates in a shear flow},}\ }\href@noop {}
  {\bibfield  {journal} {\bibinfo  {journal} {Journal of Fluid Mechanics}\ }
  (\bibinfo {year} {in press, 2019})}\BibitemShut {NoStop}%
\bibitem [{\citenamefont {Spicer}\ and\ \citenamefont
  {Pratsinis}(1996)}]{Spicer1996}%
  \BibitemOpen
  \bibfield  {author} {\bibinfo {author} {\bibfnamefont {P.~T.}\ \bibnamefont
  {Spicer}}\ and\ \bibinfo {author} {\bibfnamefont {S.~E.}\ \bibnamefont
  {Pratsinis}},\ }\bibfield  {title} {\enquote {\bibinfo {title} {Coagulation
  and fragmentation: Universal steady-state particle-size distribution},}\
  }\href@noop {} {\bibfield  {journal} {\bibinfo  {journal} {AIChE journal}\
  }\textbf {\bibinfo {volume} {42}},\ \bibinfo {pages} {1612--1620} (\bibinfo
  {year} {1996})}\BibitemShut {NoStop}%
\bibitem [{\citenamefont {Pandya}\ and\ \citenamefont
  {Spielman}(1983)}]{Pandya1983}%
  \BibitemOpen
  \bibfield  {author} {\bibinfo {author} {\bibfnamefont {J.~D.}\ \bibnamefont
  {Pandya}}\ and\ \bibinfo {author} {\bibfnamefont {L.~A.}\ \bibnamefont
  {Spielman}},\ }\bibfield  {title} {\enquote {\bibinfo {title} {Floc breakage
  in agitated suspensions: effect of agitation rate},}\ }\href@noop {}
  {\bibfield  {journal} {\bibinfo  {journal} {Chemical Engineering Science}\
  }\textbf {\bibinfo {volume} {38}} (\bibinfo {year} {1983})}\BibitemShut
  {NoStop}%
\bibitem [{\citenamefont {Ireland}\ \emph
  {et~al.}(2016{\natexlab{b}})\citenamefont {Ireland}, \citenamefont {Bragg},\
  and\ \citenamefont {Collins}}]{IrelandJFM2016}%
  \BibitemOpen
  \bibfield  {author} {\bibinfo {author} {\bibfnamefont {P.~J.}\ \bibnamefont
  {Ireland}}, \bibinfo {author} {\bibfnamefont {A.~D.}\ \bibnamefont {Bragg}},
  \ and\ \bibinfo {author} {\bibfnamefont {L.~R.}\ \bibnamefont {Collins}},\
  }\bibfield  {title} {\enquote {\bibinfo {title} {The effect of {R}eynolds
  number on inertial particle dynamics in isotropic turbulence. {P}art 1.
  simulations without gravitational effects},}\ }\href@noop {} {\bibfield
  {journal} {\bibinfo  {journal} {Journal of Fluid Mechanics}\ }\textbf
  {\bibinfo {volume} {796}},\ \bibinfo {pages} {617--658} (\bibinfo {year}
  {2016}{\natexlab{b}})}\BibitemShut {NoStop}%
\bibitem [{\citenamefont {Liu}\ and\ \citenamefont
  {Hrenya}(2018)}]{liu2018cluster}%
  \BibitemOpen
  \bibfield  {author} {\bibinfo {author} {\bibfnamefont {Peiyuan}\ \bibnamefont
  {Liu}}\ and\ \bibinfo {author} {\bibfnamefont {Christine~M}\ \bibnamefont
  {Hrenya}},\ }\bibfield  {title} {\enquote {\bibinfo {title} {Cluster-induced
  deagglomeration in dilute gravity-driven gas-solid flows of cohesive
  grains},}\ }\href@noop {} {\bibfield  {journal} {\bibinfo  {journal}
  {Physical Review Letters}\ }\textbf {\bibinfo {volume} {121}},\ \bibinfo
  {pages} {238001} (\bibinfo {year} {2018})}\BibitemShut {NoStop}%
\end{thebibliography}%

\end{document}